\documentclass[12pt]{article}
\usepackage[top=2cm,bottom=3cm,left=2cm,right=2cm]{geometry}
\usepackage{amsmath}
\usepackage{amsfonts}
\usepackage{stmaryrd}
\usepackage{bm}
\usepackage{hyperref}
\usepackage{upgreek}
\usepackage[mathscr]{euscript}
\usepackage{graphicx}
\usepackage{mathtools}

\newcommand{\inverttriangle}{%
               \mathrel{\raisebox{.1em}{%
               \reflectbox{\rotatebox[origin=c]{180}{$\triangle$}}}}}
               
\numberwithin{equation}{section}
\numberwithin{figure}{section}

\def\eq#1{(\ref{eq:#1})}
\def\lineup{\!\!\!\!\!\!\!\!&&}

\def\d{\partial}
\def\eps{\epsilon}

\def\deg{\mathrm{deg}}

\def\M{{\bf M}}
\def\m{{\bf m}}
\def\Q{{\bf Q}}
\def\n{{\bm \upeta}}

\def\b{{\bf b}}
\def\c{{\bf c}}

\def\mmu{{\bm \upmu}}

\def\D{{\bf D}}
\def\H{{\bf \hat{H}}}

\def\G{{\bf \hat{G}}}

\def\[{\big[}
\def\]{\big]}

\def\hPsiR{\widehat{\Psi}_\mathrm{R}}
\def\hPhiNS{\widehat{\Phi}_\mathrm{NS}}
\def\PsiR{\Psi_\mathrm{R}}
\def\PsiNS{\Psi_\mathrm{NS}}
\def\PhiNS{\Phi_\mathrm{NS}}
\def\PiNS{\Pi_\mathrm{NS}}
\def\PiR{\Pi_\mathrm{R}}
\def\HNres{\mathcal{H}_\mathrm{NS}^{\mathrm{restricted}}}
\def\HRres{\mathcal{H}_{\mathrm{R}}^{\mathrm{restricted}}}
\def\Htres{\widetilde{\mathcal{H}}^{\mathrm{restricted}}}
\def\Xit{\tilde{\xi }}
\def\Xxit{\tilde{\bm{\upxi}}}
\def\Xt {\widetilde{X}}
\def\Xxt{\widetilde{\bf X}}

\begin{document}

\begin{titlepage}
\rightline{\tt LMU-ASC 07/16}
\rightline{\tt UT-Komaba/16-1}

\begin{center}
\vskip 3.5cm

{\Large \bf{Complete Action for Open Superstring Field Theory}}

\vspace{.25cm}

{\Large \bf{with Cyclic $A_\infty$ Structure}}

\vskip 1.0cm

{\large {Theodore Erler$^{1}$, Yuji Okawa$^{2}$ and Tomoyuki Takezaki$^{2}$}}

\vskip 1.0cm

$^{1}${\it {Arnold Sommerfeld Center, Ludwig-Maximilians University}}\\
{\it {Theresienstrasse 37, 80333 Munich, Germany}}\\
tchovi@gmail.com\\
\vspace{.5cm}
$^{2}${\it Institute of Physics, The University of Tokyo} \\
{\it Komaba, Meguro-ku, Tokyo 153-8902, Japan}\\
okawa@hep1.c.u-tokyo.ac.jp, takezaki@hep1.c.u-tokyo.ac.jp\\

\vskip 2.0cm

{\bf Abstract}
\end{center}

We construct a gauge invariant action for the Neveu-Schwarz and Ramond sectors of open superstring field theory realizing a cyclic $A_\infty$ structure, 
providing the first complete and fully explicit solution to the classical Batalin-Vilkovisky master equation in superstring field theory. We also demonstrate the equivalence of our action to the Wess-Zumino-Witten-based construction of Kunitomo and one of the authors.

\vskip 1.0cm
\noindent 

\noindent

\end{titlepage}

\tableofcontents


\section{Introduction}

A major obstacle to the construction of superstring field theories has been formulating an action for the Ramond sector. The last few months have seen remarkable progress on this problem. In~\cite{complete} a complete action for open superstring field theory was formulated by restricting the off-shell state space of the Ramond field so as to reproduce the correct integration over the fermionic modulus in the Ramond propagator.\footnote{The construction of~\cite{complete} is based on a very old idea for formulating the free action for the Ramond string field~\cite{West1,West2,Terao,Yamaron,Kugo,Belopolsky,Kunitomo}, which with the proper understanding is equivalent to the formulation of Witten \cite{Witten}. However, the construction of \cite{complete} gives the first consistent nonlinear extension of this free action. } A closely related approach has recently been developed by Sen \cite{1PIR,SenBV}, which has a somewhat simpler worldsheet realization at the cost of introducing spurious free fields. 

We are now in a position to complete the construction of all classical superstring field theories. The construction of \cite{complete} was realized by extending the Neveu-Schwarz (NS) open superstring field theory of Berkovits \cite{Berkovits1,Berkovits2} to include the Ramond (R) sector. The Berkovits theory gives an elegant Wess-Zumino-Witten-like (WZW-like) action for the NS sector in the large Hilbert space~\cite{FMS} and is a suitable starting point for the study of tachyon condensation and classical solutions \cite{BSZ,supermarg,Oksupermarg,Okrealsupermarg,FKsuper,KOsuper,simplesupermarg,supervac}. However, the question of recent interest is how to construct other superstring field theories and how to quantize them. In this capacity the Berkovits formulation is not ideal, since it does not immediately generalize to type II closed superstrings,\footnote{Some attempts to provide a WZW-like formulation of closed type II superstring field theory are described in~\cite{Matsunaga1,Matsunaga2}. For heterotic string field theory a WZW-like formulation in the large Hilbert space is well established~\cite{heterotic}, and its extension to the Ramond sector would be interesting to consider \cite{KunHeterotic1,KunHeterotic2,KunHeterotic3}.} and, despite some attempts \cite{superBV,BerkBV,Torii1,Torii2,Torii3,superBV2}, it is not known how to properly define the gauge-fixed path integral.

For this reason, in this paper we turn our attention to a different form of open string field theory which uses the small Hilbert space and realizes a cyclic $A_\infty$ structure. The construction of superstring field theories based on $A_\infty$ and $L_\infty$ algebras is attractive since all forms of superstring field theory can in principle be described in this language \cite{Muenster,WittenSS,ClosedSS,Ramond,SenBV}. In addition, the definition of the gauge-fixed path integral is straightforward thanks to the close relation between homotopy algebras, Batalin-Vilkovisky quantization, and the Feynman-graph decomposition of moduli spaces of Riemann surfaces  (or their supergeometrical extension\footnote{The manner in which picture changing operators in the vertices implement integration over odd moduli has not yet been made fully explicit, though the computation of the four-point amplitude in \cite{INOT} has given some preliminary insight. However, it follows from the computation of the $S$-matrix \cite{Konopka} that the tree-level actions and equations of motion constructed so far correctly integrate over the supermoduli spaces of punctured disks and spheres.}) \cite{Zwiebach,SZ1,SZ2}. 

Our construction of open superstring field theory extends the NS open superstring field theory of \cite{WittenSS} to include the Ramond sector, and the interactions are built from Witten's open string star product dressed with picture changing insertions. Part of the work for constructing this theory was done in \cite{Ramond}, which gives classical equations of motion describing the interactions between the NS and R sectors. Our task is to modify the equations of motion so that they can be derived from an action. This requires, specifically, that the equations of motion realize a cyclic $A_\infty$ structure, where the notion of cyclicity is provided by the inner products defining the NS and R kinetic terms. Interestingly, the action we find for the Ramond sector turns out to be identical to that of \cite{complete} after the appropriate translation of NS degrees of freedom \cite{OkWB,WB,WBlarge}.

This paper is organized as follows. In section \ref{sec:bckd} we review the formulation of the Ramond sector kinetic term used in \cite{complete} and the NS and Ramond equations of motion described in \cite{Ramond}. In section~\ref{sec:small} we construct an action by requiring compatibility of the equations of motion with the bilinear form defining the Ramond sector kinetic term. First we describe the picture changing insertion which plays a central role in defining the vertices. Then we give an explicit discussion of the 2-string product, generalize to the higher string products, and provide a proof that the resulting $A_\infty$ structure is cyclic. We also describe how the construction can be translated into the formulation of the Ramond kinetic term used by Sen \cite{1PIR,SenBV}. In section \ref{sec:large} we relate our construction to the WZW-based formulation developed by Kunitomo and one of the authors \cite{complete}. We end with some concluding remarks.

\bigskip

\noindent{\bf Note Added:} While this paper was in preparation, we were informed of independent work by Konopka and Sachs addressing the same problem. Their work should appear concurrently \cite{SK}. See also \cite{Matsunaga3} for related discussion.

\section{Background}
\label{sec:bckd}

In this section we review the Ramond kinetic term \cite{complete} and equations of motion \cite{Ramond}. To describe compositions of string products and their interrelations in an efficient manner, we will make extensive use of the coalgebra formalism. The coalgebra formalism expresses string products in terms of {\it coderivations} or {\it cohomomorphisms} acting on the tensor algebra generated from the open string state space $\widetilde{\mathcal{H}}$:
\begin{equation}T\widetilde{\mathcal{H}} \equiv \widetilde{\mathcal{H}}^{\otimes 0}\ \oplus\ \widetilde{\mathcal{H}}\ \oplus\ \widetilde{\mathcal{H}}^{\otimes 2}\ \oplus\ \widetilde{\mathcal{H}}^{\otimes 3}\ \oplus\ ...\ \ .\end{equation}
Coderivations will be denoted in boldface, and cohomomorphisms with a ``hat" and in boldface. String fields can be described by {\it group-like elements} of the tensor algebra. The coalgebra formalism works efficiently if we use a shifted grading of the open string field called {\it degree}. The degree of a string field $A$ is defined to be its Grassmann parity $\eps(A)$ plus one:
\begin{equation}\deg(A) = \eps(A) + 1\ \ \ \mathrm{mod}\ \mathbb{Z}_2.\end{equation}
For a detailed description of all the relevant definitions, formulas, and the notational conventions, see \cite{WB}. 

\subsection{Ramond Kinetic Term} 
\label{subsec:kinetic}
 
Let us start by summarizing what is needed to have a consistent open string field theory kinetic term from the perspective of an action realizing a cyclic $A_\infty$ structure. We need three things:
\begin{description}
\item{(A)} A state space $\mathcal{H}$, perhaps a subspace of the full CFT state space, which is closed under the action of the BRST operator $Q$. The BRST cohomology at ghost number 1 computed in $\mathcal{H}$ reproduces the appropriate spectrum of open string states. 
\item{(B)} A symplectic form $\omega$ on the state space $\mathcal{H}$. This is a linear map from two copies of the state space into complex numbers,
\begin{equation}\omega:\mathcal{H}\otimes\mathcal{H}\to\mathbb{C},\end{equation}
which is graded antisymmetric,
\begin{equation}\omega(A,B) = -(-1)^{\deg(A)\deg(B)}\omega(B,A),\end{equation}
and nondegenerate. We sometimes write the symplectic form as $\langle \omega|$, and write $\omega(A,B)\equiv \langle\omega|A\otimes B$. We assume that $\omega$ is nonzero only when acting on states whose ghost number adds up to $3$.
\item{(C)} The BRST operator must be {\it cyclic} with respect to the symplectic form $\omega$:
\begin{equation}\omega(QA,B) = -(-1)^{\deg(A)}\omega(A,QB).\end{equation}
Equivalently
\begin{equation}\langle\omega|(Q\otimes\mathbb{I}+\mathbb{I}\otimes Q) = 0,\end{equation}
where $\mathbb{I}$ is the identity operator on the state space.
\end{description}
If these three criteria are met, a string field theory kinetic term can be written as
\begin{equation}S= \frac{1}{2}\omega(\Psi,Q\Psi),\end{equation}
where $\Psi$ is a degree even and ghost number 1 dynamical string field in $\mathcal{H}$. Variation of the action produces the expected equations of motion
$Q\Psi = 0$, and the action has the linearized gauge invariance 
\begin{equation}\Psi' = \Psi +Q\Lambda,\end{equation}
where $\Lambda \in \mathcal{H}$ is degree odd and carries ghost number 0.

Let us see how this story applies to the NS and R sectors of the open superstring. We consider the RNS formulation of the open superstring, described by a $c=15$ matter boundary superconformal field theory tensored with the $c=-15$ ghost boundary superconformal field theory $b,c,\beta,\gamma$. The $\beta\gamma$ system may be bosonized to the $\xi,\eta,e^\phi$ system \cite{FMS}. We will write the eta zero mode $\eta_0$ as $\eta$. The state space of the open superstring is the direct sum of an NS component $\mathcal{H}_{\mathrm{NS}}$ and a Ramond component $\mathcal{H}_{\mathrm{R}}$:
\begin{equation}\widetilde{\mathcal{H}}= \mathcal{H}_{\mathrm{NS}}\oplus\mathcal{H}_{\mathrm{R}}.\end{equation}
We use $\widetilde{\mathcal{H}}$ to denote the combined state space. Formulating the NS kinetic term requires a subspace of $\mathcal{H}_\mathrm{NS}$ 
consisting of states at picture $-1$ and in the small Hilbert space. The BRST operator preserves this subspace, and has the correct cohomology at ghost number 1. The symplectic form can be defined by the small Hilbert space BPZ inner product (up to a sign from the shifted grading):\footnote{The elementary correlator in the small Hilbert space will be normalized as $\langle c\d c\d^2 ce^{-2\phi}(0)\rangle_S=-2\times Z^{\mathrm{matter}}$, where $Z^{\mathrm{matter}}$ is the disk partition function in the matter boundary conformal field theory. In the large Hilbert space the elementary correlator will be normalized as $\langle\xi  c\d c\d^2 ce^{-2\phi}(0)\rangle_L=2\times Z^{\mathrm{matter}}$, with the opposite sign.}
\begin{equation}\omega_S(A,B) \equiv (-1)^{\deg(A)}\langle A,B\rangle_S,\end{equation}
where the subscript $S$ denotes the small Hilbert space. Furthermore, the BRST operator is cyclic with respect to $\omega_S$. Since conditions (A), (B) and (C) are met, we can write the NS kinetic term as
\begin{equation}S = \frac{1}{2}\omega_S(\PsiNS,Q\PsiNS),\end{equation}
where the dynamical NS string field $\PsiNS\in\mathcal{H}_\mathrm{NS}$ is in the small Hilbert space ($\eta\PsiNS=0$), is degree even, and carries ghost number 1 and picture $-1$. Though it is not needed to formulate the NS kinetic term, it will be useful to consider the large Hilbert space symplectic form $\omega_L$ defined in terms of the large Hilbert space BPZ inner product by
\begin{equation}\omega_L(A,B) \equiv (-1)^{\deg(A)}\langle A,B\rangle_L,\end{equation}
where the subscript $L$ denotes the large Hilbert space. 

Now let us describe the Ramond kinetic term. The major technical problem in this respect is defining an appropriate symplectic form. For this purpose we introduce two picture changing operators:
\begin{eqnarray}
\mathscr{X} \lineup \equiv  -\delta(\beta_0)G_0 + \delta'(\beta_0)b_0,\label{eq:scrX}\\
\mathscr{Y} \lineup \equiv  -c_0\delta'(\gamma_0), 
\end{eqnarray}
where $G_0$ is the zero mode of the supercurrent. The operator $\mathscr{X}$ is degree even and carries ghost number $0$ and picture $1$, and $\mathscr{Y}$ is degree even and carries ghost number $0$ and picture $-1$. Since these operators depend on $\beta\gamma$ zero modes, they only act on states in the Ramond sector. Moreover, it is clear that $\mathscr{X}$ should not act on states which are annihilated by $\beta_0$ and $\mathscr{Y}$ should not act on states which are annihilated by $\gamma_0$. For this reason we will always assume that $\mathscr{X}$ and $\mathscr{Y}$ act on states in the small Hilbert space at the following pictures: 
\begin{eqnarray}
\mathscr{X}:\lineup \ \mathrm{small\ Hilbert\ space},\ \mathrm{picture}\  -\!3/2,\nonumber\\
\mathscr{Y}:\lineup \ \mathrm{small\ Hilbert\ space},\ \mathrm{picture}\  -\!1/2. \label{eq:XYrest}
\end{eqnarray}
In particular, all pictures besides picture $-3/2$ either contain states annihilated by $\beta_0$ or are BPZ conjugate to pictures containing states annihilated by $\beta_0$. Similarly, all pictures besides picture $-1/2$ either contain states annihilated by $\gamma_0$ or are BPZ conjugate to pictures containing states annihilated by $\gamma_0$. Assuming $\mathscr{X}$ and $\mathscr{Y}$ act on the appropriate picture as above, they satisfy
\begin{equation}\mathscr{X}\mathscr{Y}\mathscr{X}= \mathscr{X},\ \ \ \ \mathscr{Y}\mathscr{X}\mathscr{Y} = \mathscr{Y},\ \ \ \ [Q,\mathscr{X}] = 0, \label{eq:preproj}\end{equation}
and are BPZ even:
\begin{equation}
\langle \omega_S| \mathscr{X}\otimes\mathbb{I} = \langle \omega_S| \mathbb{I}\otimes \mathscr{X},\ \ \ \ \langle \omega_S| \mathscr{Y}\otimes\mathbb{I}= \langle \omega_S| \mathbb{I}\otimes \mathscr{Y}.
\end{equation}
Note that \eq{preproj} implies that the operator $\mathscr{X}\mathscr{Y}$ is a projector:
\begin{equation}(\mathscr{X}\mathscr{Y})^2 = \mathscr{X}\mathscr{Y}.\end{equation}
This projector selects a subspace 
\begin{equation}\HRres\subset \mathcal{H}_\mathrm{R}\end{equation}
of Ramond states which satisfy
\begin{equation}\mathscr{X}\mathscr{Y} A = A,\ \ \ \ A\in\HRres.\end{equation}
We will call this the {\it restricted space}. To ensure that the action of $\mathscr{X}\mathscr{Y}$ is well defined, we will assume that the restricted space only contains states in the small Hilbert space and at picture $-1/2$. We claim that the restricted space allows for the definition of a Ramond kinetic term, and to see it, we check conditions (A), (B) and (C). First note that the restricted space is preserved by the action of the BRST operator:
\begin{equation}
  \mathscr{X}\mathscr{Y}QA = \mathscr{X}\mathscr{Y}Q\mathscr{X}\mathscr{Y} A =\mathscr{X}\mathscr{Y}\mathscr{X}Q\mathscr{Y}A = \mathscr{X}Q\mathscr{Y}A = Q\mathscr{X}\mathscr{Y} A = QA,\ \ \ \ A\in \HRres.
\end{equation}
Moreover, the cohomology of $Q$ computed in $\HRres$ reproduces the correct physical spectrum~\cite{coh}. Therefore condition (A) is met. Next, we define a symplectic form on $\HRres$ by
\begin{equation}\omega_S(\mathscr{Y} A,B),\ \ \ \ A,B\in\HRres.\end{equation}
Graded antisymmetry follows from the fact that $\mathscr{Y}$ is BPZ even and the fact that $\omega_S$ is graded antisymmetric. Nondegeneracy follows from the fact that $\mathscr{Y}A = 0$ implies $A = 0$ upon operating with $\mathscr{X}$, and $\omega_S$ is nondegenerate on the subspace of Ramond states at pictures $-1/2$ and $-3/2$. Therefore condition (B) is met. Finally, we have
\begin{eqnarray}
\omega_S(\mathscr{Y}A,QB) \lineup = \omega_S(\mathscr{Y}A,Q\mathscr{X}\mathscr{Y}B) \nonumber\\
\lineup = \omega_S(\mathscr{Y}A,\mathscr{X}Q\mathscr{Y}B)\nonumber\\
\lineup =\omega_S(\mathscr{X}\mathscr{Y}A,Q\mathscr{Y}B) \nonumber\\
\lineup =  \omega_S(A,Q\mathscr{Y}B)\nonumber\\
\lineup = - (-1)^{\deg(A)}\omega_S(\mathscr{Y} QA,B),\ \ \ \ \ \ A,B\in\HRres,
\end{eqnarray}
so condition (C) is met. Therefore, we can write a free action for the Ramond string field as
\begin{equation}
S= \frac{1}{2}\omega_S(\mathscr{Y}\PsiR,Q\PsiR),
\end{equation}
where the dynamical Ramond string field $\PsiR$ is in the small Hilbert space ($\eta\PsiR = 0$), is degree even, carries ghost number 1 and picture $-1/2$, and satisfies $\mathscr{X}\mathscr{Y}\PsiR = \PsiR$.

We can package the dynamical NS and R string fields together into a string field:
\begin{equation}
\widetilde{\Psi} =\PsiNS+\PsiR. 
\end{equation}
We call this the ``composite string field." It is an element of the state space
\begin{equation}\Htres = \HNres\oplus \HRres,\end{equation}
which we call the ``composite restricted space." In the NS sector, the space $\HNres$ consists of states in the small Hilbert space at picture $-1$. In the Ramond sector, the space $\HRres$ is defined as above.  We define a ``composite symplectic form"
\begin{equation}\widetilde{\omega}:\Htres\otimes\Htres \to\mathbb{C}\end{equation} 
by 
\begin{equation}\langle \widetilde{\omega}| \equiv \langle \omega_S|_0| + \langle \omega_S|_2|\mathscr{Y}\otimes\mathbb{I},\end{equation}
where, following notation to be introduced in a moment, $\langle \omega_S|_0|$ is nonzero only when contracting two NS states, and $\langle \omega_S|_2|$ is nonzero only when contracting two Ramond states. From the above discussion, it is clear that the composite restricted space together with the composite symplectic form satisfy conditions (A), (B) and (C), so we can write the kinetic term as
\begin{equation}
S = \frac{1}{2}\widetilde{\omega}(\widetilde{\Psi},Q\widetilde{\Psi}),
\end{equation}
which describes the free propagation of both the NS and R states.

\subsection{Ramond Equations of Motion}
\label{subsec:EOM}

Now that we have a free action for the NS and R sectors, our task will be to add interactions. The structure of interactions at the level of the equations of motion was described in \cite{Ramond}. It is helpful to review this before considering the action. 

The equations of motion are characterized by a sequence of degree odd multi-string products:
\begin{equation}\widetilde{M}_1\equiv Q,\ \ \widetilde{M}_2,\ \ \widetilde{M}_3,\ \ \widetilde{M}_4,\ \ ...\ .\end{equation}
We call these ``composite products" since they encapsulate the multiplication of both NS and R states. We require three properties: 
\begin{description}
\item{(I)} The composite products satisfy the relations of an $A_\infty$ algebra. Equivalently, if $\widetilde{\M}_{n+1}$ is the coderivation corresponding to $\widetilde{M}_{n+1}$, the sum 
\begin{equation}\widetilde{\M}\equiv \widetilde{\M}_1+\widetilde{\M}_2 + \widetilde{\M}_3 + \widetilde{\M}_4+...\end{equation}
defines a nilpotent coderivation on the tensor algebra:\footnote{Commutators of multi-string products are always graded with respect to degree \cite{WittenSS}. Commutators of string fields, computed with the open string star product, are graded with respect to Grassmann parity. When taking commutators of operators (or equivalently commutators of 1-string products) the degree and Grassmann gradings are equivalent.}
\begin{equation}[\widetilde{\M},\widetilde{\M}]= 0.\end{equation}
\item{(II)} The composite products are defined in the small Hilbert space. Equivalently, the coderivation $\widetilde{\M}$ commutes with the coderivation $\n$ representing the eta zero mode: 
\begin{equation}[\n,\widetilde{\M}] = 0.\end{equation}
\item{(III)} The composite products carry the required ghost and picture number so that the equations of motion,
\begin{equation}
0 = Q\widetilde{\Psi} + \widetilde{M}_2(\widetilde{\Psi},\widetilde{\Psi}) +\widetilde{M}_3(\widetilde{\Psi},\widetilde{\Psi},\widetilde{\Psi})+...,\label{eq:EOM}
\end{equation}
have an NS component at ghost number $2$ and picture $-1$, and a Ramond component at ghost number $2$ and picture $-1/2$.  
\end{description}
When we write the equations of motion, the dynamical Ramond string field does not have to be in the restricted space. Formulating the equations of motion in the restricted space is closely related to constructing the action, and will be described later. However, we still assume that $\PsiR$ is in the small Hilbert space, is degree even, and carries ghost number 1 and picture $-1/2$.
 
We will construct the composite products by placing picture changing insertions on Witten's associative star product:
\begin{equation}m_2(A,B) \equiv (-1)^{\deg(A)} A*B.\end{equation}
The generalization to other forms of open string multiplication (for example, the star product with ``stubs" \cite{ClosedSS,Ramond}) is closely related to the generalization to heterotic and type II superstring field theories, and will be left for future work. The BRST operator, the eta zero mode, and the star product satisfy
\begin{eqnarray}
0\lineup = [\Q,\Q],\ \ \ \ \, 0 = [\n,\Q],\ \ \ \ \, 0 = [\n,\n],\nonumber\\
0\lineup = [\Q,\m_2], \ \ \ 0 = [\n,\m_2],\ \ \ 0 = [\m_2,\m_2]. 
\end{eqnarray}
This says that $Q$ and $\eta$ are nilpotent and commute, that $Q$ and $\eta$ are derivations of the star product, and that the star product is associative. Equivalently, $\Q,\n$ and $\m_2$ define three mutually commuting $A_\infty$ structures. Though it is not important for the equations of motion, we note that the star product is cyclic with respect to the small (and large) Hilbert space symplectic form:
\begin{equation}\langle\omega_S|(m_2\otimes\mathbb{I}+\mathbb{I}\otimes m_2) = 0.\end{equation}
Similarly the eta zero mode is cyclic with respect to the large Hilbert space symplectic form.
 
Because $\PsiNS$ and $\PsiR$ carry different picture, the composite products $\widetilde{M}_{n+1}$ must provide a different amount of picture depending on how many NS and R states are being multiplied. To keep track of this, it will be useful to invoke the concept of {\it Ramond number}. A multi-string product has Ramond number $r$ if it is nonvanishing only when the number of Ramond inputs minus the number of Ramond outputs is equal to $r$. We will write the Ramond number of a product using a vertical slash followed by an index indicating the Ramond number. For example, $b_m|_r$ is an $m$-string product of Ramond number $r$. The definition of Ramond number implies that the product $b_m|_r$ has the property 
\begin{eqnarray}
b_m|_r\Big(\ r\ \mathrm{Ramond\ states}\ \Big) \lineup = \mathrm{NS\  state},\nonumber\\
b_m|_r\Big(\ r\!+\!1\ \mathrm{Ramond\ states}\ \Big) \lineup = \mathrm{R\  state},\nonumber\\
b_m|_r\Big(\ \mathrm{otherwise}\ \Big)\lineup = 0.
\end{eqnarray}
Any product can be written as a unique sum of products of definite Ramond number:
\begin{equation}b_m = b_m|_{-1} + b_m|_0 + b_m|_1 + ... + b_m|_m.\end{equation}
The Ramond number of $b_m$ is bounded between $-1$ and $m$ since $b_m$ can have at most $m$ Ramond inputs and at most $1$ Ramond output. Since Ramond number is conserved when composing products, it is conserved when taking commutators of coderivations:
\begin{equation}[\b_m,\c_n]|_s = \sum_{r=-1}^s [\b_m|_r,\c_n|_{s-r}], \end{equation}
with the understanding that commutators in this sum vanish if the Ramond number exceeds the number of inputs of the product. As an example of this identity, note that associativity of the star product implies
\begin{eqnarray}
0\lineup =[\m_2,\m_2]|_0 = [\m_2|_0,\m_2|_0],\label{eq:m2R0}\\
0\lineup =[\m_2,\m_2]|_2 = 2 [\m_2|_0,\m_2|_2],\label{eq:m2R2}\\
0\lineup = [\m_2,\m_2]|_4 = [\m_2|_2,\m_2|_2],
\end{eqnarray}
where the star product is broken into components of definite Ramond number as
\begin{equation}\m_2 = \m_2|_0 + \m_2|_2.\end{equation}
The components of the star product with odd Ramond number vanish identically. 

We are now ready to describe the equations of motion constructed in \cite{Ramond}. The composite products $\widetilde{M}_{n+2}$ have a component at Ramond number $0$ and a component at Ramond number $2$:
\begin{equation}\widetilde{M}_{n+2} = M_{n+2}|_0+m_{n+2}|_2,\label{eq:comp}\end{equation}
which carry the following picture and ghost numbers:
\begin{eqnarray}
M_{n+2}|_0:\lineup\ \ \mathrm{picture}\ n+1, \ \ \mathrm{ghost\ number}\ -n,\\
m_{n+2}|_2:\lineup\ \ \mathrm{picture}\ n,\ \ \ \ \ \ \ \,\! \mathrm{ghost\ number}\ -n.
\end{eqnarray}
 The 1-string product $M_1|_0$ is identified with the BRST operator 
\begin{equation}M_1|_0\equiv Q,\end{equation}
and $m_2|_2$ is the Ramond number 2 component of Witten's open string star product. We also define {\it bare products} of odd degree and {\it gauge products} of even degree:
\begin{eqnarray}
\mathrm{bare\ products}\ \ m_{n+2}|_0:\lineup\ \ \mathrm{picture}\ n,\ \ \ \ \ \ \ \mathrm{ghost\ number}\ -n,\\
\mathrm{gauge\ products}\ \ \ \mu_{n+2}|_0:\lineup\ \ \mathrm{picture}\ n+1, \ \ \mathrm{ghost\ number}\ -n-1.
\end{eqnarray}
The bare product $m_2|_0$ is the Ramond number zero component of Witten's open string star product. We define generating functions
\begin{eqnarray}
\M|_0(t) \lineup \equiv \sum_{n=0}^\infty t^n \M_{n+1}|_0,\label{eq:Mgen}\\
\m|_2(t) \lineup \equiv \sum_{n=0}^\infty t^n \m_{n+2}|_2,\\
\m|_0(t) \lineup \equiv \sum_{n=0}^{\infty} t^n \m_{n+2}|_0,\\
\mmu|_0(t) \lineup \equiv \sum_{n=0}^{\infty} t^n \mmu_{n+2}|_0,\label{eq:mpgen}
\end{eqnarray}
which are postulated to satisfy the differential equations
\begin{eqnarray}
\frac{d}{dt}\M|_0(t) \lineup = [\M|_0(t),\mmu|_0(t)],\label{eq:Mdiff}\\
\frac{d}{dt}\m|_2(t) \lineup = [\m|_2(t),\mmu|_0(t)],\label{eq:mdiff}\\
\frac{d}{dt}\m|_0(t) \lineup = [\m|_0(t),\mmu|_0(t)],\label{eq:mpdiff}\\
\ [\n,\mmu|_0(t)] \lineup = \m|_0(t).\phantom{\Big(}\label{eq:mum}
\end{eqnarray}
Expanding in powers of $t$ gives a recursive system of equations which determine higher products in terms of sums of commutators of lower ones.
A crucial step in solving this system of equations concerns \eq{mum}, which defines the gauge product $\mu_{n+2}|_0$ in terms of the bare product $m_{n+2}|_0$. The solution of \eq{mum} requires a choice of contracting homotopy of $\n$.\footnote{In this context, a contracting homotopy for $\n$ is a degree odd linear operator $\Xi\circ$ acting on the vector space of coderivations which satisfies $[\n,\Xi\circ\D] + \Xi\circ[\n,\D] = \D$ for an arbitrary coderivation $\D$. } This choice influences the configuration of picture changing insertions which appear in the vertices, and will determine whether or not the equations of motion can be derived from an action. 

The products can be usefully characterized by the cohomomorphism
\begin{equation}
\G(t) \equiv \mathcal{P}\exp\left[\int_0^t ds\, \mmu|_0(s)\right],\label{eq:Gt}
\end{equation}
where the path ordering is in sequence of increasing $s$ from left to right. In particular, the generating functions take the form
\begin{eqnarray}
\M|_0(t) \lineup = \G(t)^{-1}\Q\G(t),\phantom{\Big(}\label{eq:M0G}\\
\m|_2(t) \lineup = \G(t)^{-1}\m_2|_2\G(t),\phantom{\Big(}\label{eq:m2G}\\
\m|_0(t) \lineup = \G (t)^{-1}\m_2|_0\G(t),\phantom{\Big(}\label{eq:m0G}\\
\mmu|_0(t)\lineup = \G(t)^{-1}\frac{d}{dt}\G(t).\label{eq:mu0G}
\end{eqnarray}
Also, using \eq{mum} and \eq{m0G} it is straightforward to show that \cite{OkWB}
\begin{eqnarray}
\n \lineup = \G^{-1}(\n-\m_2|_0)\G.\label{eq:etaG}
\end{eqnarray}
Here and in what follows, all objects are evaluated at $t=1$ when the dependence on $t$ is not explicitly indicated. The coderivation representing the composite products is 
\begin{eqnarray}\widetilde{\M}\lineup = \M|_0 + \m|_2 \nonumber\\
\lineup = \G^{-1}(\Q+\m_2|_2)\G.\end{eqnarray}
From this expression it immediately follows that 
\begin{equation}[\widetilde{\M},\widetilde{\M}] = 0,\ \ \ \ [\n,\widetilde{\M}] = 0,\label{eq:Ainfsmall}\end{equation}
because $\Q,\m_2$ and $\n$ are mutually commuting $A_\infty$ structures. Therefore the composite products satisfy $A_\infty$ relations and are in the small Hilbert space. 

\section{The Action}
\label{sec:small}

Now we can bring the Ramond kinetic term and equations of motion together to define an action:
\begin{equation}
S = \frac{1}{2}\widetilde{\omega}(\widetilde{\Psi},Q\widetilde{\Psi}) + \frac{1}{3}\widetilde{\omega}(\widetilde{\Psi},\widetilde{M}_2(\widetilde{\Psi},\widetilde{\Psi}))+\frac{1}{4}\widetilde{\omega}(\widetilde{\Psi},\widetilde{M}_3(\widetilde{\Psi},\widetilde{\Psi},\widetilde{\Psi}))+...,\label{eq:action}
\end{equation}
where $\widetilde{\Psi}$ is the composite string field and $\widetilde{M}_{n+1}$ are the composite products introduced in subsection~\ref{subsec:EOM}. Since we now consider the action, the dynamical Ramond string field must belong to the restricted space.

When we vary the action, it is assumed that we should reproduce the equations of motion
\begin{equation}0=Q\widetilde{\Psi} + \widetilde{M}_2(\widetilde{\Psi},\widetilde{\Psi}) + \widetilde{M}_3(\widetilde{\Psi},\widetilde{\Psi},\widetilde{\Psi}) + ...\ .\end{equation}
However, this requires that the composite products are {\it cyclic} with respect to the composite symplectic form: 
\begin{equation}\langle\widetilde{\omega}|\big(\widetilde{M}_{n+1}\otimes\mathbb{I}+\mathbb{I}\otimes \widetilde{M}_{n+1}\big)= 0\ \ \ \ \mathrm{on}\ \ \Htres.\end{equation}
Thus the composite products define a cyclic $A_\infty$ algebra. Cyclicity does not follow automatically from the construction of the equations of motion given in subsection \ref{subsec:EOM}, but requires a special choice of picture changing insertions inside the vertices. More technically, it requires a special choice of contracting homotopy for $\n$ in the solution of \eq{mum}, and our task is to find it. 

\subsection{Picture Changing Insertion}
\label{subsec:PCO}

The picture changing insertions in the action are defined with the operator
\begin{equation}
\Xit: \mathrm{degree\ odd},\ \ \mathrm{ghost\ number}\  -\!1,\ \ \mathrm{picture}\  1, 
\end{equation}
which has the following properties:
\begin{description}
\item{\ \ \ \ \  1)} $\Xit$ is a contracting homotopy for $\eta$:\ \  $[\eta,\Xit] = 1$,

\item{\ \ \ \ \  2)} $\Xit  $ is BPZ even:\ \  $\langle \omega_L|\Xit \otimes\mathbb{I} = \langle\omega_L|\mathbb{I}\otimes\Xit $,

\item{\ \ \ \ \  3)} $[Q,\Xit  ] = \mathscr{X}$ when acting on a Ramond state at picture $-3/2$ in the small Hilbert space,

\item{\ \ \ \ \  4)} $\Xit^2=0$.
\end{description}
Property 1) is needed to define a contracting homotopy for $\n$ in the solution of \eq{mum}. Properties~2) and 3) will be needed in the proof of cyclicity. Property 4) will not be essential for our purposes, but we would like to have it anyway. 

A natural candidate for $\Xit  $ is the operator $\Theta(\beta_0)$ as used in \cite{complete}, which in particular satisfies
\begin{equation}[Q,\Theta(\beta_0)]=\mathscr{X}.\end{equation}
However, we must be careful to avoid acting $\Theta(\beta_0)$ on states annihilated by $\beta_0$. This means that $\Theta(\beta_0)$ can only act ``safely" on the states:
\begin{equation}\Theta(\beta_0):\ \mathrm{small\ Hilbert\ space,\ picture}\, -\! 3/2.\label{eq:Thsm}\end{equation}
It may seem somewhat unnatural to require that $\Theta(\beta_0)$ acts on the small Hilbert space, since generically it maps into the large Hilbert space. Let us explain why this is necessary. Suppose $\Theta(\beta_0)$ could act on an arbitrary state $A$ at picture $-3/2$ in the large Hilbert space. Then we should be able to contract with a state $B$ at picture $-1/2$,
\begin{equation}\langle \Theta(\beta_0)A,B\rangle_L,\end{equation}
and obtain a finite result. Now suppose $A=QA'$ and $B'=QB$. Then using the BPZ even property of $\mathscr{X}$ gives
\begin{equation}
\langle \Theta(\beta_0)A,B\rangle_L= \langle A',\mathscr{X} B\rangle_L+(-1)^{\eps(A')+1} \langle \Theta(\beta_0)A',B'\rangle_L.
\end{equation}
We have assumed that the left hand side is finite, and the second term on the right hand side should be finite by the same assumption. However, this contradicts the fact that the first term on the right hand side can be infinite if $B$ is annihilated by $\beta_0$. Therefore, the action of $\Theta(\beta_0)$ in the large Hilbert space must generally be singular.

This causes problems with a direct attempt to identify $\Theta(\beta_0)$ with the operator $\Xit   $. Nevertheless, it was shown in \cite{complete} that $\Theta(\beta_0)$ at least formally satisfies properties $1)-4)$. However, in \cite{complete} it was assumed that $\Theta(\beta_0)$ never acts on states annihilated by $\beta_0$. Here we would like to provide a setting where this assumption is justified. First, note that \eq{Thsm} implies that we can define operators $\Theta(\beta_0)\eta$ and $\eta\Theta(\beta_0)$ acting on the following states: 
\begin{eqnarray}
\lineup \Theta(\beta_0)\eta:\ \mathrm{large\ Hilbert\ space,\ picture}\, -\! 1/2,\nonumber\\
\lineup \eta\Theta(\beta_0):\ \mathrm{large\ Hilbert\ space,\ picture}\, -\! 1/2.
\end{eqnarray}
The operator $\Theta(\beta_0)\eta$ is well defined since $\eta$ maps from the large Hilbert space at picture $-1/2$ into the small Hilbert space at picture $-3/2$, after which we can act with $\Theta(\beta_0)$. The operator $\eta\Theta(\beta_0)$ is defined by BPZ conjugation of $\Theta(\beta_0)\eta$. Therefore we have 
\begin{equation}\langle \omega_L| \eta\Theta(\beta_0)\otimes\mathbb{I} = \langle\omega_L|\mathbb{I}\otimes \Theta(\beta_0)\eta\end{equation}
when acting on states in the large Hilbert space at picture $-1/2$. We also have 
\begin{equation}\eta\Theta(\beta_0) + \Theta(\beta_0)\eta = 1\label{eq:Thid1}\end{equation}
when acting in the large Hilbert space at picture $-1/2$. We can also say that $\Theta(\beta_0)$ is nilpotent in the sense that 
\begin{equation}\eta\Theta(\beta_0)^2\eta = 0,\label{eq:Thid3}\end{equation}
which similarly holds on states in the large Hilbert space at picture $-1/2$.

Having understood the limitations of $\Theta(\beta_0)$, we can search for a more acceptable alternative. For this purpose we introduce the operator \cite{INOT,WittenSS}
\begin{equation}\xi  \equiv\oint_{|z|=1}\frac{dz}{2\pi i} f(z)\xi (z),\end{equation}
where the function $f(z)$ is holomorphic in the vicinity of the unit circle. The function $f(z)$ can be chosen so that $\xi $ is BPZ even and commutes with $\eta$ to give 1:
\begin{equation}
\langle \omega_L| \xi \otimes\mathbb{I} = \langle \omega_L|\mathbb{I}\otimes\xi ,\ \ \ \ [\eta,\xi ] = 1.
\end{equation}
In addition $\xi ^2=0$. Therefore $\xi $ realizes properties 1), 2) and 4), but it does not realize property~3). Rather, the BRST variation gives the operator
\begin{equation}
X\equiv [Q,\xi ],
\label{eq:X}
\end{equation}
which is not the same as $\mathscr{X}$. This can be fixed by defining a ``hybrid" operator between $\xi $ and $\Theta(\beta_0)$: 
\begin{equation}
\Xit    \equiv \xi  + (\Theta(\beta_0)\eta\xi  - \xi ) P_{-3/2} + (\xi \eta\Theta(\beta_0) -\xi )P_{-1/2},
\end{equation}
where $P_{n}$ projects onto states at picture $n$. Note that $\Theta(\beta_0)$ always appears here in allowed combinations with $\eta$ acting on allowed pictures. Note also that $\Xit$ reduces to $\xi $ when acting on NS states, as is appropriate for defining the NS superstring field theory \cite{WittenSS}.
It is also clear that $\Xit$ is BPZ even, and so realizes property 2). To see that property 3) is realized, let us define the picture changing operator
\begin{equation}\Xt \equiv[Q,\Xit].\end{equation}
Note that in general $\Xt$ is different from $\mathscr{X}$ defined in \eq{scrX} and $X$ defined in \eq{X}. However, $\Xt$ is identical to $\mathscr{X}$ when it acts on a state $A$ in the small Hilbert space at picture $-3/2$:
\begin{eqnarray}
\Xt  A\lineup  = [Q,\Theta(\beta_0)\eta\xi ] A \nonumber\\
\lineup = \Big(\mathscr{X}\eta\xi  + \Theta(\beta_0)\eta X\Big) A\nonumber\\
\lineup =  \Big(\mathscr{X}[\eta,\xi ] + \Theta(\beta_0)[\eta, X]\Big) A\nonumber\\
\lineup = \mathscr{X} A,
\end{eqnarray}
so property 3) is realized. Now let us confirm properties 1) and 4). Note
\begin{equation}P_{n} \eta = \eta P_{n+1},\end{equation}
and compute
\begin{eqnarray}
[\eta,\Xit] \lineup = 1+ \eta\Big(\Theta(\beta_0)\eta\xi  - \xi \Big) P_{-3/2}\nonumber\\
\lineup \ \ \ \ \ + \Big[\eta\Big(\xi \eta\Theta(\beta_0) -\xi \Big)+\Big(\Theta(\beta_0)\eta\xi  - \xi \Big)\eta\Big]P_{-1/2}\nonumber\\
\lineup\ \ \ \ \ +\Big(\xi \eta\Theta(\beta_0) -\xi \Big)\eta P_{1/2}\nonumber\\
\lineup = 1+ (\eta\xi  - \eta\xi ) P_{-3/2}\nonumber\\
\lineup \ \ \ \ \ + \Big[\eta\Theta(\beta_0) -\eta\xi +\Theta(\beta_0)\eta - \xi \eta\Big]P_{-1/2}\nonumber\\
\lineup\ \ \ \ \ +(\xi \eta -\xi \eta) P_{1/2}\nonumber\\
\lineup = 1,
\end{eqnarray}
where we used \eq{Thid1} and $[\eta,\xi ]=1$. Finally let us check property 4):
\begin{eqnarray}
\Xit   ^2 \lineup = \xi ^2 + \Big(\Theta(\beta_0)\eta\xi  - \xi \Big) P_{-3/2}\Big(\Theta(\beta_0)\eta\xi  - \xi \Big) P_{-3/2}\nonumber\\
\lineup\ \ \ \ \ \ \, +\, \Big(\xi \eta\Theta(\beta_0) -\xi \Big)P_{-1/2}\Big(\xi \eta\Theta(\beta_0) -\xi \Big)P_{-1/2}\nonumber\\
\lineup\ \ \ \ \ \ \, +\, \xi  \Big(\Theta(\beta_0)\eta\xi  - \xi \Big) P_{-3/2}+ \Big(\Theta(\beta_0)\eta\xi  - \xi \Big) P_{-3/2}\xi \nonumber\\
\lineup\ \ \ \ \ \ \, +\, \xi \Big(\xi \eta\Theta(\beta_0) -\xi \Big)P_{-1/2}\!+\!\Big(\xi \eta\Theta(\beta_0) -\xi \Big)P_{-1/2}\xi  \nonumber\\
\lineup\ \ \ \ \ \ \,+\,\Big(\Theta(\beta_0)\eta\xi  - \xi \Big) P_{-3/2}\Big(\xi \eta\Theta(\beta_0) -\xi \Big)P_{-1/2}\nonumber\\
\lineup\ \ \ \ \ \ \,+\,\Big(\xi \eta\Theta(\beta_0) -\xi \Big)P_{-1/2}\Big(\Theta(\beta_0)\eta\xi  - \xi \Big) P_{-3/2}\nonumber\\
\lineup = \xi ^2 +\xi  \Big(\Theta(\beta_0)\eta\xi  - \xi \Big) P_{-3/2}+ \Big(\Theta(\beta_0)\eta\xi  - \xi \Big)\xi  P_{-5/2} + \xi \Big(\xi \eta\Theta(\beta_0) -\xi \Big)P_{-1/2}\nonumber\\
\lineup\ \ \ \ \ \ \,+\,\Big(\xi \eta\Theta(\beta_0) -\xi \Big)\xi  P_{-3/2} +\Big(\xi \eta\Theta(\beta_0) -\xi \Big)\Big(\Theta(\beta_0)\eta\xi  - \xi \Big) P_{-3/2}.
\end{eqnarray}
In the second step we commuted all projectors to the right and dropped terms with a pair of projections into incompatible pictures. Using $\xi ^2=0$ this further simplifies
\begin{eqnarray}
\Xit ^2 \lineup =\xi  \Theta(\beta_0)\eta\xi   P_{-3/2} +\xi \eta\Theta(\beta_0)\xi  P_{-3/2} 
+\Big(\xi \eta\Theta(\beta_0)^2\eta\xi  -\xi \eta\Theta(\beta_0)\xi -\xi \Theta(\beta_0)\eta\xi \Big)P_{-3/2}\nonumber\\
\lineup = \xi \eta\Theta(\beta_0)^2\eta\xi  P_{-3/2}\nonumber\\
\lineup = 0,
\end{eqnarray}
which vanishes as a consequence of \eq{Thid3}. Therefore we have a definition of the picture changing insertion $\Xit$ with all necessary properties.

It is worth mentioning that $\mathscr{X}$ and $\Theta(\beta_0)$ cannot be expressed in an elementary way in terms of the local picture changing insertions $X(z)$ and $\xi(z)$. Therefore, the computation of correlation functions with $\mathscr{X}$ and $\Theta(\beta_0)$ does not appear to be straightforward. However, a recipe for computations with such operators was given in \cite{revisited} in the context of $\beta\gamma$ correlation functions, where they may be represented as formal integrals
\begin{equation}
\mathscr{X} \equiv \int d\zeta \int d\tilde{\zeta}\, e^{\zeta G_0 -\tilde{\zeta}\beta_0},\ \ \ \ \Theta(\beta_0) \equiv -\int d\tilde{\zeta}\,\frac{e^{-\tilde{\zeta}\beta_0}}{\tilde{\zeta}},\label{eq:intrep}
\end{equation}
where $\zeta$ is an odd integration variable and $\tilde{\zeta}$ is an even integration variable. The key point is that the integral over the even variable $\tilde{\zeta}$ should be understood algebraically, analogous to the Berezin integral over the odd variable $\zeta$, rather than as an ordinary integral in the sense of analysis. One difficulty, however, is the appearance of a singular factor $\tilde{\zeta}^{-1}$ in the integral for $\Theta(\beta_0)$. This is related to the fact that $\Theta(\beta_0)$ is an operator in the large Hilbert space, and therefore its precise definition must go slightly beyond the formalism of \cite{revisited}. Here we give one prescription for dealing with this. We may express $\Theta(\beta_0)$ in the form
\begin{equation}
\Theta(\beta_0) = \xi_0 +\Delta,
\end{equation}
where
\begin{equation}\Delta \equiv \Theta(\beta_0) - \xi_0,\end{equation}
and $\xi_0$ is the zero mode of the $\xi$ ghost. The term $\Delta$ can be represented as an algebraic integral 
\begin{eqnarray}
\Delta = -\int d\tilde{\zeta}\,\frac{e^{-\tilde{\zeta}\beta_0}}{\tilde{\zeta}} + \oint_{|z|=1} \frac{dz}{2\pi i}\frac{1}{z}\int d\tilde{\zeta}\,\frac{e^{-\tilde{\zeta}\beta(z)}}{\tilde{\zeta}}.
\end{eqnarray}
Since the first term is independent of $z$, we can write $\Delta$ as 
\begin{equation}
\Delta = \oint_{|z|=1} \frac{dz}{2\pi i}\frac{1}{z} \int d\tilde{\zeta}\,\frac{1}{\tilde{\zeta}}\Big(-e^{-\tilde{\zeta}\beta_0} + e^{-\tilde{\zeta}\beta(z)}\Big).
\end{equation}
Finally, we represent the integrand as the integral of a total derivative,
\begin{equation}
\Delta = \oint_{|z|=1} \frac{dz}{2\pi i}\frac{1}{z} \int d\tilde{\zeta}\,\frac{1}{\tilde{\zeta}}\int_0^1 dt\,\frac{d}{dt}e^{-\tilde{\zeta}(t\beta(z)+(1-t)\beta_0)},
\end{equation}
and taking the derivative with respect to $t$ gives
\begin{eqnarray}
\Delta = \int_0^1 dt \oint_{|z|=1} \frac{dz}{2\pi i}\frac{1}{z} \int d\tilde{\zeta}(\beta_0-\beta(z)) e^{-\tilde{\zeta}(t\beta(z)+(1-t)\beta_0)}.
\end{eqnarray}
Note that the problematic factor $\tilde{\zeta}^{-1}$ is canceled. The upshot is that we have defined $\Theta(\beta_0)$ as a sum of $\xi_0$, which can be understood in the bosonized $\beta\gamma$ system, and $\Delta$, which can be evaluated following \cite{revisited}. To see how this definition can be applied, note that the computation of a typical open string field theory vertex requires evaluating correlation functions with multiple insertions of $\Theta(\beta_0)$:
\begin{equation}
\Theta^{(1)}\Theta^{(2)}...\Theta^{(n)},
\end{equation}
where $\Theta^{(i)}$ represent appropriate conformal transformations of $\Theta(\beta_0)$. Writing $\Theta(\beta_0) = \xi_0+\Delta$ produces cross terms of the form
\begin{equation}
\xi^{(1)}\xi^{(2)}...\,\xi^{(m)}\Delta^{(m+1)}\Delta^{(m+2)}...\,\Delta^{(n)},
\end{equation}
where $\xi^{(i)}$ and $\Delta^{(i)}$ represent appropriate conformal transformations of $\xi_0$ and $\Delta$, respectively. Since $(\xi^{(1)})^2 = 0$, we can replace these insertions with 
\begin{equation}
\xi^{(1)}(\xi^{(2)}-\xi^{(1)})...\,(\xi^{(m)}-\xi^{(1)})\Delta^{(m+1)}\Delta^{(m+2)}...\,\Delta^{(n)}.
\end{equation}
We can now drop the factor $\xi^{(1)}$, which only serves to saturate the $\xi$ zero mode in the large Hilbert space, and evaluate the remaining factors using $\beta\gamma$ correlation functions as in \cite{revisited}.

An important question is whether our choice of picture changing insertions $\Xit$ and $\Xt$ avoid contact divergences in vertices and amplitudes, as appear for example when we use a local picture changing insertion in the cubic vertex \cite{Wendt}. In the NS sector such divergences are absent since the picture changing insertions appear as holomorphic contour integrals \cite{INOT,WittenSS}. In the Ramond sector, the picture changing insertions appear as $\Theta(\beta_0)$ and $\mathscr{X}$; to our knowledge, such operators can only be divergent in the presence of a zero mode of the path integral associated with $\beta_0$. We have taken some care to ensure that $\Theta(\beta_0)$ and $\mathscr{X}$ operate on states of pictures where such zero modes are absent, and therefore the vertices are expected to be finite. Explicit calculations with similar operators will be discussed in upcoming work \cite{OO}, and no contact divergences appear.

\subsection{The 2-String Product}
\label{subsec:2string}

We are ready to construct the products defining the action. Let us start by expanding the equations of motion out to second order in the string field and in NS and R components:
\begin{eqnarray}
0\lineup =Q\PsiNS + M_2|_0(\PsiNS,\PsiNS) + m_2|_2(\PsiR,\PsiR)+\ ... \ ,\\
0\lineup =Q\PsiR + M_2|_0(\PsiNS,\PsiR)+M_2|_0(\PsiR,\PsiNS)+\ ...\ .\label{eq:REOM2}
\end{eqnarray}
In \cite{WittenSS} the product of two NS states was defined by
\begin{equation}
M_2|_0 = \frac{1}{3}\Big(Xm_2|_0 + m_2|_0(X\otimes \mathbb{I}+ \mathbb{I}\otimes X)\Big) \ \ \ \ \ \ \ \ (\mathrm{multiplying\ NS\ states}).\label{eq:NSM20}
\end{equation}
This definition does not work for multiplying an NS and R state, since it does not multiply into the restricted space in the Ramond sector. For this reason we take
\begin{equation}
M_2|_0 = \mathscr{X} m_2|_0 \ \ \ \ \ \ \ \ \ \ \ \ \ \ \ \ \ \ \ \ \ (\mathrm{multiplying\ NS\ and\ R\ state\ in}\ \Htres).
\end{equation}
Because $\mathscr{X}\mathscr{Y}\mathscr{X} = \mathscr{X}$, this product satisfies $\mathscr{X}\mathscr{Y}M_2|_0 = M_2|_0$ and therefore maps into the restricted space. Note that this definition of $M_2|_0$ differs from \cite{Ramond}, where it was assumed that $M_2|_0$ multiplies two NS states and an NS and R state in the same way.  To make notation uniform it is helpful to write $X$ and $\mathscr{X}$ together using the picture changing operator $\Xt$, so we define 
\begin{equation}
M_2|_0 \equiv 
\left\{\begin{matrix*}[l] 
& {\displaystyle \frac{1}{3}}\Big(\Xt m_2|_0 + m_2|_0(\Xt\otimes \mathbb{I}+ \mathbb{I}\otimes \Xt)\Big) &\ \ \ (0\ \mathrm{Ramond\ inputs}) \phantom{\Bigg]}\\
& \Xt m_2|_0 &\ \ \ (1\ \mathrm{Ramond\ input})\phantom{\Bigg]}
\end{matrix*}\right..\label{eq:M202}
\end{equation}
The full composite 2-product is then
\begin{equation}
\widetilde{M}_2 \equiv 
\left\{\begin{matrix*}[l] 
& {\displaystyle \frac{1}{3}}\Big(\Xt m_2|_0 + m_2|_0(\Xt\otimes \mathbb{I}+ \mathbb{I}\otimes \Xt)\Big) &\ \ \ (0\ \mathrm{Ramond\ inputs}) \phantom{\Bigg]}\\
& \Xt m_2|_0 &\ \ \ (1\ \mathrm{Ramond\ input})\phantom{\Bigg]}\\
&  m_2|_2 &\ \ \ (2\ \mathrm{Ramond\ inputs})\phantom{\Bigg]}
\end{matrix*}\right..\label{eq:subsection}
\end{equation}
Note that using $\widetilde{X}$ gives a definition of the product $M_2|_0$ between arbitrary states in $\widetilde{\mathcal{H}}$. Following the discussion of subsection \ref{subsec:EOM}, the product $M_2|_0$ should be derived from a gauge 2-product $\mu_2|_0$ and bare 2-product $m_2|_0$ satisfying the formulas
\begin{eqnarray}
\M_2|_0 \lineup = [\Q,\mmu_2|_0],\\
\ [\n,\mmu_2|_0]\lineup = \m_2|_0. 
\end{eqnarray}
The last equation defines $\mu_2|_0$ in terms of $m_2|_0$ with an appropriate choice of contracting homotopy for $\n$. The choice of contracting homotopy which produces our preferred definition of $M_2|_0$ is realized by the following gauge 2-product:
\begin{equation}
\mu_2|_0 \equiv 
\left\{\begin{matrix*}[l] 
& {\displaystyle \frac{1}{3}}\Big(\Xit m_2|_0 - m_2|_0(\Xit\otimes \mathbb{I}+ \mathbb{I}\otimes \Xit)\Big) &\ \ \ (0\ \mathrm{Ramond\ inputs}) \phantom{\Bigg]}\\
& \Xit m_2|_0 &\ \ \ (1\ \mathrm{Ramond\ input})\phantom{\Bigg]}
\end{matrix*}\right..\label{eq:mu20}
\end{equation}
This completes the definition of the equations of motion up to second order.

Now we want to see that the equations of motion can be derived from an action.  This requires that  the composite 2-product is cyclic:
\begin{equation}\langle\widetilde{\omega}| \mathbb{I}\otimes\widetilde{M}_2=-\langle\widetilde{\omega}| \widetilde{M}_2\otimes\mathbb{I} \ \ \ \mathrm{on}\ \Htres.\end{equation}
Note that cyclicity only needs to hold when the vertex is evaluated on the composite restricted space, since this is the space of the dynamical string field appearing in the action. Outside this space the products will not be cyclic, and in fact the notion of cyclicity itself is somewhat problematic since $\mathscr{Y}$ may act on a state of the wrong picture. The demonstration of cyclicity goes slightly differently depending on the arrangement of NS and R states in the vertex. Let us discuss for example the case  
\begin{equation}\langle\widetilde{\omega}| (\mathbb{I}\otimes\widetilde{M}_2) (R_1\otimes R_2 \otimes N_1),\end{equation}
where $R_1,R_2$ are Ramond states and $N_1$ is an NS state in $\Htres$. Expanding into components of definite Ramond number, we have
\begin{eqnarray}
\langle\widetilde{\omega}| (\mathbb{I}\otimes\widetilde{M}_2) (R_1\otimes R_2 \otimes N_1) \lineup = \Big(\langle\omega_S|_0| + \langle \omega_S|_2|\mathscr{Y}\otimes\mathbb{I}\Big) \Big(\mathbb{I}\otimes(M_2|_0+m_2|_2)\Big) (R_1\otimes R_2 \otimes N_1)\nonumber\\
\lineup = \langle \omega_S|_2|(\mathscr{Y} \otimes M_2|_0)(R_1\otimes R_2\otimes N_1).
\end{eqnarray}
The product $m_2|_2$ drops out since it does not multiply a sufficient number of Ramond states, and $\langle \omega_S|_0|$ drops out since it contracts too many Ramond states. Plugging in \eq{M202} we obtain
\begin{eqnarray}
\langle\widetilde{\omega}| (\mathbb{I}\otimes\widetilde{M}_2) (R_1\otimes R_2 \otimes N_1) \lineup =  \langle \omega_S|_2|(\mathscr{Y} \otimes \Xt m_2|_0 )(R_1\otimes R_2\otimes N_1)\nonumber\\
\lineup =  \langle \omega_S|_2|(\mathscr{Y} \otimes \mathscr{X}m_2|_0 )(R_1\otimes R_2\otimes N_1)\nonumber\\
\lineup = \langle \omega_S|_2|(\mathscr{X}\mathscr{Y} \otimes m_2|_0 )(R_1\otimes R_2\otimes N_1)\nonumber\\
\lineup = \langle \omega_S|_2|(\mathbb{I} \otimes m_2|_0 )(R_1\otimes R_2\otimes N_1)\nonumber\\
\lineup = \langle \omega_S|(\mathbb{I} \otimes m_2 )(R_1\otimes R_2\otimes N_1).
\end{eqnarray}
In the second step we noted that $\Xt $ acts on a state of picture $-3/2$ in the small Hilbert space, and therefore can be replaced by $\mathscr{X}$. 
In the third step we used that $\mathscr{X}$ is BPZ even and in the fourth step we used the fact that $R_1$ is in the restricted space. Finally we dropped the Ramond number labels since in this context they are redundant. Note that in these steps it is important to assume that the states are in $\Htres$. Next consider 
\begin{eqnarray}
\langle\widetilde{\omega}| (\widetilde{M}_2\otimes\mathbb{I}) (R_1\otimes R_2 \otimes N_1) \lineup = \Big(\langle \omega_S|_0|+  \langle \omega_S|_2|\mathscr{Y}\otimes\mathbb{I}\Big)\Big((M_2|_0+m_2|_2)\otimes \mathbb{I}\Big)(R_1\otimes R_2\otimes N_1)\nonumber\\
\lineup = \langle \omega_S|_0|(m_2|_2\otimes \mathbb{I})(R_1\otimes R_2\otimes N_1)\nonumber\\
\lineup = \langle \omega_S|(m_2\otimes \mathbb{I})(R_1\otimes R_2\otimes N_1).
\end{eqnarray}
We therefore have
\begin{equation}
\langle\widetilde{\omega}| (\widetilde{M}_2\otimes\mathbb{I}+\mathbb{I}\otimes\widetilde{M}_2) (R_1\otimes R_2 \otimes N_1)
= \langle \omega_S|(m_2\otimes \mathbb{I}+\mathbb{I}\otimes m_2)(R_1\otimes R_2\otimes N_1) = 0,
\end{equation}
which vanishes because the open string star product is cyclic. The proof of cyclicity for the other combinations $R_1\otimes N_1\otimes R_2$ and $N_1\otimes R_1\otimes R_2$ goes similarly. When all inputs are NS states, cyclicity follows from the construction of the NS open superstring field theory in \cite{WittenSS}. Therefore we have a cubic vertex consistent with a cyclic $A_\infty$ structure. 

\subsection{Higher Products}

Now let us discuss the generalization to higher string products. Defining the higher products requires a choice of contracting homotopy for $\n$ in the solution of the equation 
\begin{equation}[\n,\mmu_{n+2}|_0]=\m_{n+2}|_0.\end{equation}
The contracting homotopy we choose defines the gauge products as follows: 
\begin{equation}
\mu_{n+2}|_0 \equiv 
\left\{\begin{matrix*}[l] 
& {\displaystyle \frac{1}{n+3}}\Big(\Xit    m_{n+2}|_0  - m_{n+2}|_0(\Xit    \otimes \mathbb{I}^{\otimes n+1}+...+ \mathbb{I}^{\otimes n+1}\otimes \Xit   )\Big) &\ \ \ (0\ \mathrm{Ramond\ inputs}) \phantom{\Bigg]}\\
& \Xit m_{n+2}|_0 &\ \ \ (1\ \mathrm{Ramond\ input})\phantom{\Bigg]}
\end{matrix*}\right..\label{eq:mun02}
\end{equation}
It is not immediately obvious that this leads to a cyclic $A_\infty$ structure. We will prove that it does in the next subsection. For now, we demonstrate two important properties, which follow from this definition: 
\begin{eqnarray}
M_{n+2}|_0 \lineup = \Xt  m_{n+2}|_0\ \ \ \  (1\ \mathrm{Ramond\ input}),\phantom{\Big]}\label{eq:Xm}\\
m_{n+2}|_2 \lineup = 0\ \ \ \ \ \ \ \ \ \ \ \ \ \ \, (3\ \mathrm{Ramond\ inputs}).\phantom{\Big]}\label{eq:mp0}
\end{eqnarray}
The first equation generalizes \eq{M202}, and implies that the interactions are consistent with the projection onto the restricted space in the Ramond sector. The second equation addresses a puzzle raised in \cite{Ramond} concerning the existence of cubic terms in the Ramond string field in the equations of motion. The existence of such terms is consistent with $A_\infty$ relations, but is not compatible with the existence of an action since the equations of motion do not possess quartic terms in the Ramond string field. (Recall that $\widetilde{M}_n$ has no component with Ramond number 4.) Therefore, the fact that $m_{n+2}|_2$ vanishes with three Ramond inputs is expected and in fact necessary to derive the equations of motion from an action. In total, then, we find that the composite products appear as follows:
\begin{equation}
\widetilde{M}_{n+2} = 
\left\{\begin{matrix*}[l] 
& \displaystyle{\frac{1}{n+3}}\Big(\Xt m_{n+2}|_0+m_{n+2}|_0(\Xt\!\otimes\!\mathbb{I}^{\otimes n+1}+...+ \mathbb{I}^{\otimes n+1}\!\otimes\! \Xt)\Big) 
& (0\ \mathrm{Ramond\ inputs}) \phantom{\bigg]}\\
& \Xt m_{n+2}|_0 & (1\ \mathrm{Ramond\ input})\phantom{\bigg]}\\
&  m_{n+2}|_2 & (2\ \mathrm{Ramond\ inputs})\phantom{\bigg]}\\
& 0 & (\mathrm{otherwise})\phantom{\bigg]}
\end{matrix*}\right. .
\end{equation}
The products $m_{n+2}|_0$ and $m_{n+2}|_2$ above are determined recursively by solving \eq{mdiff} and \eq{mpdiff} 
with our choice of gauge products \eq{mun02}.

To streamline the proof of \eq{Xm} and \eq{mp0}, it will be useful to introduce the projection operator
\begin{equation}\pi_n^{r}: T\widetilde{\mathcal{H}}\to T\widetilde{\mathcal{H}},\label{eq:pinr}\end{equation}
which selects $n$-string states containing $r$ Ramond factors (and therefore $n-r$ NS factors). This projector commutes in a simple way through coderivations of products with definite Ramond number:
\begin{eqnarray}
\pi_{m+1}^r \, \b_{n}|_s \lineup = \b_{n}|_s\, \pi_{m+n}^{s+r}.\label{eq:codpi}
\end{eqnarray}
We also define
\begin{equation}\pi_n = \sum_{r=0}^n \pi_n^r,\end{equation}
which projects onto $n$-string states with an undetermined number of Ramond factors. With these projectors we can express \eq{Xm} and \eq{mp0} in a more useful form using coderivations. First we write
\begin{eqnarray}
\M_{n+2}|_0\pi_{n+2}^1 \lineup = \Xxt  \m_{n+2}|_0\pi_{n+2}^1,\\
\m_{n+2}|_2\pi_{n+2}^3 \lineup = 0,
\end{eqnarray}
where $\Xxt$ is the coderivation corresponding to $\Xt$. Commuting the projectors through the coderivations using \eq{codpi} gives
\begin{eqnarray}
\pi_1^1\M_{n+2}|_0 \lineup = \Xt  \pi_1^1 \m_{n+2}|_0 ,\\
\pi_1^1 \m_{n+2}|_2 \lineup = 0.
\end{eqnarray}
Summing over $n$ then implies
\begin{eqnarray}
\pi_1^1 (\M|_0 - \Q) \lineup = \Xt \pi_1^1 \m|_0\phantom{\Big[},\label{eq:Xm2}\\
\pi_1^1 \m|_2 \lineup = 0\phantom{\Big]}.\label{eq:mp02}
\end{eqnarray}
In the first equation we subtract $Q$ since \eq{Xm2} only applies to the 2-string product and higher. 

To prove \eq{Xm2} and \eq{mp02} it is helpful to first derive the form of $\G^{-1}$ when it produces one Ramond output:
\begin{equation}\pi_1^1 \G^{-1}.\end{equation}
To compute this, note that  
\begin{eqnarray}
\frac{d}{dt}\Big[\pi_1^1\G(t)^{-1}\Big] = -\pi_1^1\mmu|_0(t)\G(t)^{-1}
\end{eqnarray}
from the definition of the path ordered exponential. Our choice of contracting homotopy for $\n$ in the Ramond sector \eq{mun02} implies
\begin{equation}\pi_1^1\mmu|_0(t) = \pi_1^1 \Xxit \m|_0(t),\end{equation}
where $\Xxit$ is the coderivation corresponding to $\Xit$. Plugging in gives
\begin{eqnarray}
\frac{d}{dt}\Big[\pi_1^1 \G(t)^{-1}\Big]\lineup = -\pi_1^1\Xxit \m|_0(t) \G(t)^{-1} = -\pi_1^1 \Xxit \G(t)^{-1}\m_2|_0,
\end{eqnarray}
where we used $\m|_0(t) = \G(t)^{-1}\m_2|_0\G(t)$. Therefore we obtain
\begin{equation}
\frac{d}{dt}\Big[\pi_1^1\G(t)^{-1}\Big]  = -\Xit \Big[\pi_1^1 \G(t)^{-1}\Big]\m_2|_0. \label{eq:diffRG}
\end{equation}
The solution is subject to the initial condition  $\G(0)^{-1} = \mathbb{I}_{T\widetilde{\mathcal{H}}}$, where 
$\mathbb{I}_{T\widetilde{\mathcal{H}}}$ is the identity operator on the tensor algebra. This determines the solution to be 
\begin{equation}\pi_1^1\G(t)^{-1} = \pi_1^1\Big[\mathbb{I}_{T\widetilde{\mathcal{H}}} - t\Xxit \m_2|_0\Big].\end{equation}
This satisfies \eq{diffRG} since $(\m_2|_0)^2 = 0$ by \eq{m2R0}. Setting $t=1$ we have
\begin{equation}\pi_1^1\G^{-1} = \pi_1^1\Big[\mathbb{I}_{T\widetilde{\mathcal{H}}} - \Xxit \m_2|_0\Big].\label{eq:GinvR}\end{equation}
This identity will play a central role in the following analysis, as it is the basis for our proof of cyclicity and the relations \eq{Xm2} and \eq{mp02}, and it provides a crucial link to the WZW-based theory in section~\ref{sec:large}. Note that expanding the path ordered exponential \eq{Gt} and integrating over the parameter in the generating function gives a general expression for $\pi_1^1\G^{-1}$:
\begin{equation}
\pi_1^1\G^{-1} = \pi_1^1\left(\mathbb{I}_{T\widetilde{\mathcal{H}}} - \mmu_2|_0 -\frac{1}{2}\mmu_3|_0 +\frac{1}{2}\mmu_2|_0\mmu_2|_0+\ ...\ \right).
\label{eq:exGinvR1}
\end{equation}
This is substantially more elaborate than \eq{GinvR}. With our choice of contracting homotopy for $\n$, the higher order products in $\pi_1^1\G^{-1}$ drop out, giving a closed form expression. 

Now we are ready to prove \eq{Xm2} and \eq{mp02}. First note that the bare products with one Ramond output simplify to  
\begin{eqnarray}\pi_1^1\m|_0  \lineup = \pi_1^1\G^{-1} \m_2|_0\G\nonumber\\
\lineup = \pi_1^1\m_2|_0 \G, \label{eq:m1R}\end{eqnarray}
since the second term in \eq{GinvR} cancels by associativity of the star product. Now consider $\M|_0$ with one Ramond output:
\begin{eqnarray}
\pi_1^1 \M|_0 \lineup = \pi_1^1 \G^{-1}\Q\G \nonumber\\
\lineup = \pi_1^1 \Big[\mathbb{I}_{T\widetilde{\mathcal{H}}} - \Xxit \m_2|_0\Big] \Q \G \nonumber\\
\lineup = \pi_1^1 \Big[\Q\G - \Q\Xxit \m_2|_0 \G +\Xxt \m_2|_0\G \Big]\nonumber\\
\lineup = \pi_1^1 \Q \Big[\mathbb{I}_{T\widetilde{\mathcal{H}}} - \Xxit \m_2|_0\Big]\G +\Xt \pi_1^1 \m_2|_0 \G \nonumber\\
\lineup = \pi_1^1\Q  +\Xt\pi_1^1\m_2|_0 \G.
\end{eqnarray}
From this we conclude
\begin{equation}\pi_1^1(\M|_0-\Q) = \Xt \pi_1^1\m|_0,\label{eq:Xmc}\end{equation}
establishing \eq{Xm}. Next consider
\begin{eqnarray}
\pi_1^1 \m|_2 \lineup = \pi_1^1\G^{-1}\m_2|_2 \G \nonumber\\
\lineup = \pi_1^1\Big[\mathbb{I}_{T\widetilde{\mathcal{H}}} - \Xxit \m_2|_0\Big]\m_2|_2 \G \nonumber\\
\lineup = \pi_1^1 \m_2|_2 \G + \Xit\pi_1^1 \m_2|_2 \m_2|_0\G,
\end{eqnarray}
where in the third line we used
\begin{equation}\m_2|_0\m_2|_2 = -\m_2|_2\m_2|_0\end{equation}
from \eq{m2R2}. Now note
\begin{equation}\pi_1^1\m_2|_2 = m_2|_2\pi_2^3 = 0.\end{equation}
This holds because the 2-string component of the state space cannot have three Ramond factors. Therefore 
\begin{equation}\pi_1^1\m|_2 = 0,\label{eq:mp0c}\end{equation}
which establishes \eq{mp0}.

\subsection{Proof of Cyclicity}
\label{subsec:cyclic}

Having constructed the products, we are ready to demonstrate cyclicity: 
\begin{equation}\langle \widetilde{\omega}|(\widetilde{M}_{n+1}\otimes\mathbb{I}+ \mathbb{I}\otimes\widetilde{M}_{n+1}) = 0\ \ \ \ \mathrm{on}\ \ \Htres.
\end{equation}
We will need to simplify this equation somewhat before we arrive at the key property responsible for cyclicity and provide its proof. Note that the cyclicity of $\widetilde{M}_1=Q$ was already demonstrated in subsection \ref{subsec:kinetic}. When the vertex acts only on NS states, cyclicity follows from the construction of the NS open superstring field theory in \cite{WittenSS}. When the vertex acts on one or three Ramond states, it vanishes identically since the symplectic form and composite products do not carry odd Ramond number. When the vertex acts on four or more Ramond states, it vanishes identically since the composite products vanish when multiplying three or more Ramond states. Therefore, all that we need to show is that the vertex is cyclic when it acts on two Ramond states:
\begin{equation}\langle \widetilde{\omega}|(\widetilde{M}_{n+2}\otimes\mathbb{I}+ \mathbb{I}\otimes\widetilde{M}_{n+2})\pi^2_{n+3} = 0\ \ \ \ \mathrm{on}\ \ \Htres
.\end{equation}
Expanding $\widetilde{M}_{n+2}$ into components of definite Ramond number, this reads
\begin{eqnarray}
\lineup \!\!\!\!\!\!\!\!\!\langle \omega_S| (m_{n+2}|_2 \otimes\mathbb{I}+\mathbb{I}\otimes m_{n+2}|_2)\pi_{n+3}^2 \nonumber\\
\lineup \ \ \ \ \ \ \ \ \ \ \ \ \ \ \ \ \ \ \ \ \ \ 
+\langle \omega_S|(\mathscr{Y}\otimes\mathbb{I}) (M_{n+2}|_0 \otimes\mathbb{I}+\mathbb{I}\otimes M_{n+2}|_0)\pi_{n+3}^2 = 0
\ \ \ \ \mathrm{on}\ \ \Htres.\label{eq:cycC1}
\end{eqnarray}
In the first term, both Ramond states must be channeled into the input of $m_{n+2}|_2$. In the second term, the Ramond states split between the input of $M_{n+2}|_0$ and the symplectic form. This means that we can simplify the second term using \eq{Xm}: 
\begin{eqnarray}
\langle \omega_S|(\mathscr{Y}\otimes\mathbb{I})(M_{n+2}|_0\otimes\mathbb{I})\pi_{n+3}^2 \lineup =
\langle \omega_S|(\mathscr{Y}\widetilde{X}m_{n+2}|_0\otimes\mathbb{I})\pi_{n+3}^2\nonumber\\
\lineup = \langle \omega_S|(\mathscr{Y}\mathscr{X}m_{n+2}|_0\otimes\mathbb{I})\pi_{n+3}^2\nonumber\\
\lineup = \langle \omega_S|(m_{n+2}|_0\otimes\mathscr{X}\mathscr{Y})\pi_{n+3}^2\nonumber\\
\lineup = \langle \omega_S|(m_{n+2}|_0\otimes\mathbb{I})\pi_{n+3}^2\ \ \ \ \mathrm{on}\ \ \Htres.\ \ \ \ 
\end{eqnarray}
In the first step we used \eq{Xm}; in the second step we used the fact that $\widetilde{X} = \mathscr{X}$ when acting on a state in the small Hilbert space at picture $-3/2$; in the third step we used that $\mathscr{X}$ and $\mathscr{Y}$ are BPZ even; in the fourth step we used that $\mathscr{X}\mathscr{Y}=1$ when acting on states in the restricted space. Then the statement of cyclicity reduces to 
\begin{equation}
\langle \omega_S| \Big((m_{n+2}|_0 +m_{n+2}|_2)\otimes\mathbb{I}+\mathbb{I}\otimes (m_{n+2}|_0+m_{n+2}|_2)\Big)\pi_{n+3}^2=0
\ \ \ \ \mathrm{on}\ \ \Htres.\label{eq:cycC2}
\end{equation}
Therefore $m_{n+2}|_0+m_{n+2}|_2$ should be cyclic with respect to the small Hilbert space symplectic form when the vertex acts on $\Htres$ including two Ramond states. Actually, we wish to make a slightly stronger hypothesis: $m_{n+2}|_0+m_{n+2}|_2$ is cyclic with respect to the {\it large} Hilbert space symplectic when the vertex acts on the {\it large} Hilbert space including two Ramond states: 
\begin{equation}
\langle \omega_L| \Big((m_{n+2}|_0 +m_{n+2}|_2)\otimes\mathbb{I}+\mathbb{I}\otimes (m_{n+2}|_0+m_{n+2}|_2)\Big)\pi_{n+3}^2=0.\label{eq:cycC3}
\end{equation}
This relation is the nontrivial property required for the proof of cyclicity. We will provide a demonstration in a moment, but first let us explain why \eq{cycC3} implies \eq{cycC2}. The small and large Hilbert space symplectic forms can be related by
\begin{equation}\langle \omega_S| = \langle\omega_L|\xi\otimes\mathbb{I},\end{equation}
where $\xi$ satisfies $[\eta,\xi]=1$. The precise form of $\xi$ is not important since its only role is to saturate the $\xi$ zero mode in the large Hilbert space CFT correlator. The left hand side of \eq{cycC2} can be expressed as
\begin{eqnarray}
\lineup \langle \omega_S| \Big((m_{n+2}|_0 +m_{n+2}|_2)\otimes\mathbb{I}+\mathbb{I}\otimes (m_{n+2}|_0+m_{n+2}|_2)\Big)\pi_{n+3}^2 \nonumber\\
\lineup \ \ \ \ \ \ =\langle \omega_L| (\xi\otimes\mathbb{I})\Big((m_{n+2}|_0 +m_{n+2}|_2)\otimes\mathbb{I}+\mathbb{I}\otimes (m_{n+2}|_0+m_{n+2}|_2)\Big)\pi_{n+3}^2,
\end{eqnarray}
where for current purposes we assume that this equation acts on the small Hilbert space, including states in $\Htres$. Now in front of $\pi_{n+3}^2$ insert the identity operator in the form
\begin{equation}\mathbb{I}^{\otimes n+3} = \eta\xi\otimes\mathbb{I}^{\otimes n+2},\end{equation}
where $\eta\xi$ is equivalent to the identity since it acts on a state in the small Hilbert space. Moving $\eta$ to the left it will commute with $\xi$ to give $1$ and otherwise act on states in the small Hilbert space to give zero. Thus we have
\begin{eqnarray}
\lineup \langle \omega_S| \Big((m_{n+2}|_0 +m_{n+2}|_2)\otimes\mathbb{I}+\mathbb{I}\otimes (m_{n+2}|_0+m_{n+2}|_2)\Big)\pi_{n+3}^2 \nonumber\\
\lineup \ \ \ \ \ \ =-\langle \omega_L| \Big((m_{n+2}|_0 +m_{n+2}|_2)\otimes\mathbb{I}+\mathbb{I}\otimes (m_{n+2}|_0+m_{n+2}|_2)\Big)\pi_{n+3}^2 (\xi\otimes\mathbb{I}^{\otimes n+2}).
\end{eqnarray}
From this we can see that \eq{cycC3} implies \eq{cycC2} when operating on $\Htres$.

We can proceed to prove \eq{cycC3} using the recursive definition of the products. However, the proof in this form requires consideration of several different cases depending on the arrangement of NS and R inputs on the left hand side of \eq{cycC3}. Earlier we encountered similar inconvenience in the proof of cyclicity of $\widetilde{M}_2$ at the end of subsection \ref{subsec:2string}. A more efficient route to the proof uses the coalgebra formalism, and therefore it is useful to review how the cyclicity is described in this language. An $n$-string product $D_n$ is cyclic with respect to a symplectic form $\omega$ if
\begin{equation}\langle \omega| (D_n\otimes\mathbb{I} +\mathbb{I}\otimes D_n)= 0.\end{equation}
If we have a sequence of cyclic $n$-string products $D_0,D_1,D_2,...$ of the same degree, the corresponding coderivation $\D = \D_0+\D_1+\D_2+...$ will satisfy
\begin{equation}
\langle \omega|\pi_2 \D = 0. \label{eq:codcyc}
\end{equation}
We then say that the coderivation $\D$ is {\it cyclic} with respect to the symplectic form $\omega$. A cohomomorphism $\H$ is {\it cyclic} with respect to $\omega$ if it satisfies
\begin{equation}\langle \omega|\pi_2 \H = \langle\omega|\pi_2. \label{eq:cohcyc}\end{equation}
An example of a cyclic cohomomorphism is
\begin{equation}\H = \mathcal{P}\exp\left[\int_0^1 ds\, {\bf h}(s)\right],\end{equation}
where ${\bf h}(s)$ are a one-parameter family of degree even cyclic coderivations. To prove that $\H$ in this form is cyclic, consider $\H(u)$ obtained by replacing the lower limit $s=0$ in the path ordered exponential above with $s=u$. Taking the derivative with respect to $u$ we find
\begin{equation}\frac{d}{du} \langle \omega|\pi_2 \H(u) = \langle \omega|\pi_2 {\bf h}(u) \H(u) = 0. \end{equation}
This vanishes on the assumption that ${\bf h}(s)$ is cyclic. Therefore, the object $\langle \omega|\pi_2 \H(u)$ is independent of $u$. Setting $u=0$ and $u=1$ reproduces \eq{cohcyc}. The construction of the NS superstring field theory~\cite{WittenSS} implies that the gauge products are cyclic with respect to the large Hilbert space symplectic form when acting on NS states. Therefore we have 
\begin{eqnarray}
\langle \omega_L|\pi_2^0 \mmu|_0(t) = 0.
\end{eqnarray}
Using \eq{Gt} this also implies
\begin{equation}
\langle \omega_L|\pi_2^0 \G = \langle \omega_L|\pi_2^0.\label{eq:Gcyc}
\end{equation}
Therefore $\G$ is cyclic in the large Hilbert space when acting on NS states.

Next it is helpful to recall a few things about the ``triangle formalism" of the product and coproduct introduced in \cite{WB}. For this purpose we will need to think about ``tensor products" of tensor algebras, which we denote with the symbol $\otimes'$ to avoid confusion with the tensor product~$\otimes$ defining $T\widetilde{\mathcal{H}}$. The product $\inverttriangle$ is a linear map from two copies of $T\widetilde{\mathcal{H}}$ into $T\widetilde{\mathcal{H}}$:
\begin{equation}\inverttriangle: T\widetilde{\mathcal{H}}\otimes'T\widetilde{\mathcal{H}} \to T\widetilde{\mathcal{H}},\end{equation}
and the coproduct $\triangle$ is a linear map from one copy of $T\widetilde{\mathcal{H}}$ into two copies of $T\widetilde{\mathcal{H}}$:
\begin{equation}\triangle: T\widetilde{\mathcal{H}} \to T\widetilde{\mathcal{H}}\otimes'T\widetilde{\mathcal{H}}.\end{equation}
The coproduct is defined by its action on tensor products of states:
\begin{equation}
\triangle A_1\otimes ... \otimes A_n = \sum_{k=0}^n (A_1\otimes ...\otimes A_k) \otimes' (A_{k+1}\otimes...\otimes A_n),
\end{equation}
where at the extremes of summation $\otimes'$ multiplies the identity of the tensor product $1_{T\widetilde{\mathcal{H}}}$. The product $\inverttriangle$ acts by replacing the primed tensor product $\otimes'$ with the tensor product $\otimes$. A coderivation $\D$ and a cohomomorphism $\H$ satisfy the following compatibility conditions with respect to the coproduct:
\begin{eqnarray}
\triangle \D \lineup = (\D\otimes' \mathbb{I}_{T\widetilde{\mathcal{H}}} + \mathbb{I}_{T\widetilde{\mathcal{H}}}\otimes'\D)\triangle,\label{eq:cod}\\
\triangle \H\lineup = (\H\otimes'\H)\triangle.\label{eq:coh}
\end{eqnarray}
These are in fact the defining properties of coderivations and cohomomorphisms. The useful identity for our computations is 
\begin{equation}\pi_{m+n} = \inverttriangle\!\! \Big[\pi_m\otimes'\pi_n\Big]\triangle.\end{equation}
A generalization which also accounts for a projection onto $r$ Ramond factors is
\begin{equation}\pi_{m+n}^r = \sum_{k=0}^r \inverttriangle\!\! \Big[\pi_m^{r-k}\otimes'\pi_n^k\Big]\triangle,\end{equation}
with the understanding that $\pi_n^k$ vanishes if $k>n$. 

Using coalgebra notation, the key equation \eq{cycC3} can be expressed as follows:
\begin{equation}\langle \omega_L| \big(\pi_2^2\m|_0+\pi_2^0 \m|_2\big)=0.\label{eq:cyc3}\end{equation}
To prove this, consider the second term on the left hand side: 
\begin{eqnarray}\langle \omega_L|\pi_2^0 \m|_2 \lineup = \langle \omega_L|\pi_2^0 \G^{-1}\m_2|_2\G\nonumber\\
\lineup = \langle \omega_L|\pi_2^0 \G\G^{-1}\m_2|_2\G\nonumber\\
\lineup = \langle \omega_L|\pi_2^0\m_2|_2\G.
\end{eqnarray}
In the second step we used the fact that $\G$ is cyclic with respect to $\omega_L$ when it only has NS outputs, as in \eq{Gcyc}. Next consider the first term of \eq{cyc3}. Expressing $\pi_2^2$ in terms of the product and coproduct gives
\begin{eqnarray}
\langle \omega_L|\pi_2^2 \m|_0 \lineup = \langle \omega_L|\!\!\inverttriangle\!\!\Big[\pi_1^1\otimes' \pi_1^1\Big]\triangle\, \m|_0\nonumber\\
\lineup = \langle\omega_L|\!\!\inverttriangle\!\!\Bigg[(\pi_1^1\otimes'\pi_1^1)(\m|_0\otimes'\mathbb{I}_{T\widetilde{\mathcal{H}}} + \mathbb{I}_{T\widetilde{\mathcal{H}}}\otimes'\m|_0)\Bigg]\triangle\nonumber\\
\lineup = \langle\omega_L|\!\!\inverttriangle\!\!\Bigg[\Big(\pi_1^1\m|_0\Big)\otimes' \pi_1^1 + \pi_1^1\otimes'\Big(\pi_1^1\m|_0\Big)\Bigg]\triangle.
\end{eqnarray}
The form of $\m|_0$ with one Ramond output is given in \eq{m1R}. Plugging in gives
\begin{equation}\langle \omega_L|\pi_2^2 \m|_0 = \langle\omega_L|\!\!\inverttriangle\!\!\Big[(\pi_1^1\m_2|_0\G)\otimes' \pi_1^1 + \pi_1^1\otimes'(\pi_1^1\m_2|_0\G)\Big]\triangle.\end{equation}
The factor $\pi_1^1$ in the two terms above can be written as
\begin{equation}
\pi_1^1 = \pi_1^1 \G^{-1}\G = \pi_1^1\G - \Xit\pi_1^1\m_2|_0\G,
\end{equation}
where we used \eq{GinvR}. Therefore we have
\begin{eqnarray}
\langle \omega_L|\pi_2^2 \m|_0 \lineup = \langle\omega_L|\!\!\inverttriangle\!\!\Bigg[\big(\pi_1^1\m_2|_0\G\big)\otimes' \big(\pi_1^1\G\big) + \big(\pi_1^1\G\big)\otimes'\big(\pi_1^1\m_2|_0\G\big)\Bigg]\triangle\nonumber\\
\lineup\ \ \ -
\langle\omega_L|\!\!\inverttriangle\!\!\Bigg[\big(\pi_1^1\m_2|_0\G\big)\otimes' \big(\Xit\pi_1^1\m_2|_0\G\big) + \big(\Xit\pi_1^1\m_2|_0\G\big)\otimes'\big(\pi_1^1\m_2|_0\G\big)\Bigg]\triangle.
\end{eqnarray}
The second term above can be simplified as follows:
\begin{eqnarray}
\lineup 
\langle\omega_L|\!\!\inverttriangle\!\!\Bigg[\Big(\mathbb{I}\otimes'\Xit\Big)\Big(\big(\pi_1^1\m_2|_0\G\big)\otimes' \big(\pi_1^1\m_2|_0\G\big)\Big) - \Big(\Xit\otimes'\mathbb{I}\Big)\Big(\big(\pi_1^1\m_2|_0\G\big)\otimes'\big(\pi_1^1\m_2|_0\G\big)\Big)\Bigg]\triangle\nonumber\\
\lineup\ \ \ \ \ \ \ \ \ \ \ \ \ \ \ \ 
=\langle\omega_L|(\mathbb{I}\otimes\Xit-\Xit\otimes\mathbb{I})\!\!\inverttriangle\!\!\Bigg[\big(\pi_1^1\m_2|_0\G\big)\otimes' \big(\pi_1^1\m_2|_0\G\big) \Bigg]\triangle,
\end{eqnarray}
which vanishes since $\Xit$ is BPZ even. With what is left we can disentangle the product and coproduct:
\begin{eqnarray}
\langle \omega_L|\pi_2^2 \m|_0 \lineup = \langle\omega_L|\!\!\inverttriangle\!\!\Bigg[(\pi_1^1\otimes'\pi_1^1)(\m_2|_0\otimes'\mathbb{I}_{T\widetilde{\mathcal{H}}}+\mathbb{I}_{T\widetilde{\mathcal{H}}}\otimes'\m_2|_0)(\G\otimes'\G)\Bigg]\triangle\nonumber\\
\lineup = \langle\omega_L|\!\!\inverttriangle\!\!\Bigg[(\pi_1^1\otimes'\pi_1^1)(\m_2|_0\otimes'\mathbb{I}_{T\widetilde{\mathcal{H}}}+\mathbb{I}_{T\widetilde{\mathcal{H}}}\otimes'\m_2|_0)\Bigg]\triangle \G\nonumber\\
\lineup = \langle\omega_L|\!\!\inverttriangle\!\!\Big[\pi_1^1\otimes'\pi_1^1\Big]\triangle \m_2|_0 \G\nonumber\\
\lineup = \langle\omega_L|\pi_2^2 \m_2|_0\G.
\end{eqnarray}
Bringing the first and second terms in \eq{cyc3} together therefore gives 
\begin{equation}
\langle \omega_L| \big(\pi_2^2\m|_0+\pi_2^0 \m|_2\big) = \langle \omega_L|(\pi_2^2 \m_2|_0+\pi_2^0\m_2|_2)\G.
\end{equation}
Commuting the projectors past the star product in the two terms gives
\begin{eqnarray}
\langle \omega_L| \big(\pi_2^2\m|_0+\pi_2^0 \m|_2\big) \lineup = \langle \omega_L|\pi_2(\m_2|_0+\m_2|_2)\pi_3^2\G\nonumber\\
\lineup = \langle\omega_L|\pi_2\m_2\pi_3^2\G\nonumber\\
\lineup = 0,
\end{eqnarray}
which vanishes since the star product is cyclic with respect to the large Hilbert space symplectic form. This completes the proof of cyclicity.

\subsection{Relation to Sen's Formulation}
\label{subsec:Sen}

Here we would like to spell out the relation between our treatment of the Ramond sector and the approach developed by Sen \cite{1PIR,SenBV}. The main advantage of Sen's approach is that it utilizes simpler picture changing insertions, which may facilitate calculations. On the other hand, the theory propagates spurious free fields and does not directly display a cyclic $A_\infty$ structure. 

Sen's approach requires two dynamical string fields
\begin{eqnarray}
\widetilde{\Psi}\lineup = \PsiNS+\PsiR,\\
\widetilde{\Pi}\lineup = \PiNS+\PiR.
\end{eqnarray}
The NS fields $\PsiNS$ and $\PiNS$ are in the small Hilbert space, degree even, and carry ghost number 1 and picture $-1$.  The Ramond fields $\PsiR$ and $\PiR$ are in the small Hilbert space, degree even, and at ghost number 1, but carry different pictures: $\PsiR$ carries picture $-1/2$ and $\PiR$ carries picture $-3/2$. In this approach it is not necessary to assume that $\mathscr{X}\mathscr{Y}\PsiR = \PsiR$. The action takes the form
\begin{equation}
S = -\frac{1}{2}\omega_S(\widetilde{\Pi},\mathcal{G}Q\widetilde{\Pi}) + \omega_S(\widetilde{\Pi},Q\widetilde{\Psi}) + \frac{1}{3}\omega_S(\widetilde{\Psi},\widetilde{b}_2(\widetilde{\Psi},\widetilde{\Psi})) + \frac{1}{4}\omega_S(\widetilde{\Psi},\widetilde{b}_3(\widetilde{\Psi},\widetilde{\Psi},\widetilde{\Psi}))+...,
\end{equation}
where $\widetilde{b}_{n+2}$ are degree odd multi-string products which appropriately multiply NS and R states, and the operator $\mathcal{G}$ is defined by
\begin{eqnarray}
\mathcal{G} \lineup = \mathbb{I}\ \ \ \ \ (\mathrm{acting\ on\ NS\ state}),\nonumber\\
\mathcal{G} \lineup = X\ \ \ \ (\mathrm{acting\ on\ R\ state}).
\end{eqnarray}
For present purposes we can assume that the picture changing operator $X$ is defined as in \eq{X}. In particular, $\mathcal{G}$ is BPZ even and $[Q,\mathcal{G}] = 0$.

The action does not realize a cyclic $A_\infty$ structure in the standard sense, but the products $\widetilde{b}_{n+2}$ satisfy a hierarchy of closely related algebraic identities. To describe them, we introduce a sequence of degree odd multi-string products
\begin{equation}
\widetilde{M}_1\equiv Q,\ \ \widetilde{M}_2,\ \ \widetilde{M}_3,\ \ \widetilde{M}_4,\ \  ...,
\end{equation}
where 
\begin{equation}\widetilde{M}_{n+2} \equiv \mathcal{G}\widetilde{b}_{n+2}\ \ \ \ \ (n=0,1,2,...).\label{eq:bc}\end{equation}
The relation to the composite products introduced earlier will be clear in a moment. The first few algebraic relations satisfied by the multi-string products are 
\begin{eqnarray}
0\lineup = Q\widetilde{b}_2(A,B) + \widetilde{b}_2(QA,B) +(-1)^{\deg(A)}\widetilde{b}_2(A,QB),\\
0\lineup = Q\widetilde{b}_3(A,B,C) + \widetilde{b}_3(QA,B,C)+(-1)^{\deg(A)}\widetilde{b}_3(A,QB,C)+(-1)^{\deg(A)+\deg(B)}\widetilde{b}_3(A,B,QC)\ \ \ \ \ \ \nonumber\\
\lineup\ \ +\widetilde{b}_2(\widetilde{M}_2(A,B),C) + (-1)^{\deg(A)}\widetilde{b}_2(A,\widetilde{M}_2(B,C)),\\
\lineup\vdots\ \ .\nonumber
\end{eqnarray}
More abstractly, the full set of algebraic relations can be described using the coderivations
\begin{eqnarray}
\widetilde{\b}\lineup \equiv \widetilde{\b}_2 + \widetilde{\b}_3 +\widetilde{\b}_4 + ...,\\
\widetilde{\M}\lineup \equiv \Q + \widetilde{\M}_2 + \widetilde{\M}_3 +\widetilde{\M}_4 + ...,
\end{eqnarray}
as
\begin{equation}\pi_1(\Q\widetilde{\b} + \widetilde{\b}\widetilde{\M}) = 0.\label{eq:bcid} \end{equation}
In addition, gauge invariance requires that the products $\widetilde{b}_{n+2}$ are cyclic with respect to the small Hilbert space symplectic form:
\begin{equation}\langle \omega_S|\pi_2\widetilde{\b} = 0.\end{equation}
Note that \eq{bc} implies
\begin{equation}
\mathcal{G}\pi_1\widetilde{\b} = \pi_1(\widetilde{\M}-\Q).
\end{equation}
Multiplying \eq{bcid} by $\mathcal{G}$ gives
\begin{eqnarray}
0 \lineup = \mathcal{G}\pi_1(\Q\widetilde{\b} + \widetilde{\b}\widetilde{\M})\nonumber\\
 \lineup =\pi_1\Big(\Q(\widetilde{\M}-\Q) + (\widetilde{\M}-\Q)\widetilde{\M}\Big)\nonumber\\
 \lineup = \pi_1\widetilde{\M}^2, 
\end{eqnarray}
which implies that the products $\widetilde{M}_{n+1}$ satisfy $A_\infty$ relations:
\begin{equation}
[\widetilde{\M},\widetilde{\M}] = 0.
\end{equation}
However, the products $\widetilde{M}_{n+1}$ are not required to be cyclic. Rather, cyclicity is realized by the products $\widetilde{b}_{n+2}$ which appear in the action. We will explain why this formulation leads to a gauge invariant action in appendix \ref{app:Sen}.

As suggested by the notation, it is natural to identify $\widetilde{M}_{n+1}$ with the composite products constructed earlier. Indeed the composite products can be written in the form
\begin{equation}
\widetilde{M}_{n+2} = \widetilde{\mathcal{G}}\ \widetilde{b}_{n+2}\ \ \ \ \ (n=0,1,2,...)\label{eq:bct}
\end{equation}
for some products $\widetilde{b}_{n+2}$, where
\begin{eqnarray}
\widetilde{\mathcal{G}} \lineup = \mathbb{I}\ \ \ \ \ (\mathrm{acting\ on\ NS\ state}),\nonumber\\
\widetilde{\mathcal{G}} \lineup = \Xt\ \ \ \ (\mathrm{acting\ on\ R\ state}).
\end{eqnarray}
This differs from \eq{bc} only by the substitution of $\widetilde{X}$ with $X$. Therefore it is natural to construct the products as before but replacing the picture changing insertion in \eq{mun02} as 
\begin{equation}\widetilde{\xi}\to\xi.\end{equation}
Then the composite products satisfy \eq{bc}, where $\widetilde{b}_{n+2}$ takes the form
\begin{eqnarray}
\widetilde{b}_{n+2}=
\left\{\begin{matrix*}[l]   \ \ \displaystyle{\frac{1}{n+3}}\Big(X m_{n+2}|_0+m_{n+2}|_0(X\!\otimes\!\mathbb{I}^{\otimes n+1}+...+ \mathbb{I}^{\otimes n+1}\!\otimes\! X)\Big) & (\mathrm{0\ Ramond\ inputs}) \phantom{\bigg[}\\
 \ \ m_{n+2}|_0 &  (\mathrm{1\ Ramond\ input}) \phantom{\bigg[}\\
 \ \ m_{n+2}|_2 &  (\mathrm{2\ Ramond\ inputs})\phantom{\bigg[} \\
 \ \ 0&   (\mathrm{otherwise})\phantom{\bigg[}
 \end{matrix*}\right. \nonumber\\  
\end{eqnarray}
with the understanding that $m_{n+2}|_0$ and $m_{n+2}|_2$ are constructed out of $\xi$ rather than $\Xit$. We can show that $\widetilde{b}_{n+2}$ satisfies \eq{bcid} by pulling a factor of $\mathcal{G}$ out of the $A_\infty$ relations for $\widetilde{M}_{n+2}$.\footnote{Note that the products $\widetilde{M}_{n+1}$ satisfy $A_\infty$ relations regardless of whether or not $X$ has a kernel. This can only be true if \eq{bcid} holds regardless of whether $X$ has a kernel. However, it is not difficult to check \eq{bcid} directly.} Furthermore, the cyclicity of $\widetilde{b}_{n+2}$ follows from the proof of \eq{cycC3} in the previous subsection with the replacement of $\Xit$ with~$\xi$.

\section{Relation to the WZW-based Formulation}
\label{sec:large}

In this section we explain the relation between our construction to the WZW-based formulation of \cite{complete}. The relation between the NS sectors was considered in \cite{OkWB,WB,WBlarge}, and our task will be to extend this analysis to the Ramond sector.

The WZW-based theory uses an NS dynamical field
\begin{equation}\hPhiNS,\end{equation} 
which is Grassmann even, carries ghost and picture number zero, and lives in the large Hilbert space (generically $\eta\hPhiNS\neq0$). The dynamical Ramond field 
\begin{equation}\hPsiR,\end{equation}
is the same kind of state as the Ramond field $\PsiR$ from the $A_\infty$ theory; it is Grassmann odd, carries ghost number 1 and picture $-1/2$, and lives in the restricted space in the Ramond sector. We will always denote objects in the WZW-based theory with a ``hat" to distinguish from corresponding objects defined in the $A_\infty$ theory. To write the NS sector of the action in WZW-like form, we introduce a one-parameter family of NS string fields $\hPhiNS(t),t\in[0,1]$, subject to the boundary conditions
\begin{equation}\hPhiNS(0) = 0,\ \ \ \ \hPhiNS(1) = \hPhiNS.\end{equation} 
The WZW-based action of \cite{complete} can be written as
\begin{equation}\widehat{S} = \frac{1}{2}\langle \mathscr{Y}\hPsiR,Q\hPsiR\rangle_S - \int_0^1dt\, \langle \widehat{A}_t(t), Q \widehat{A}_\eta(t)+ 
(\widehat{F}(t)\hPsiR)^2\rangle_L.\label{eq:lHSaction}
\end{equation}
The ``potentials" are defined by
\begin{eqnarray}
\widehat{A}_\eta(t)\lineup \equiv (\eta e^{\hPhiNS(t)})e^{-\hPhiNS(t)},\nonumber\\
\widehat{A}_t(t)\lineup\equiv\left(\frac{d}{dt} e^{\hPhiNS(t)}\right) e^{-\hPhiNS(t)}.
\end{eqnarray}
The object $\widehat{F}(t)$ is a linear operator acting on string fields, defined by
\begin{equation}
\widehat{F}(t) \equiv \frac{1}{\mathbb{I}-\Xit \mathrm{ad}_{\widehat{A}_\eta(t)}},
\end{equation}
where $\mathrm{ad}_{\widehat{A}_\eta(t)}$ refers to the adjoint action of $\widehat{A}_\eta(t)$:
\begin{equation}\mathrm{ad}_{\widehat{A}_\eta(t)} \Psi \equiv [\widehat{A}_\eta(t),\Psi].\end{equation}
All products of string fields are computed with the open string star product $AB=A*B$, and all commutators of string fields are graded with respect to Grassmann parity. The WZW-based action only depends on the value of $\hPhiNS(t)$ at $t=1$. Variation of the action produces the equations of motion \cite{complete}
\begin{eqnarray}
0\lineup = Q \widehat{A}_\eta + (\widehat{F}\hPsiR)^2,\label{eq:lNSEOM}\\
0\lineup = Q \widehat{F}\hPsiR. \label{eq:lREOM}
\end{eqnarray}
Unless the dependence on $t$ is explicitly indicated, we will assume $t=1$ here and in what follows.

\subsection{Field Redefinition}

The relation between these string field theories can be extracted by inspection of the equations of motion \cite{WB}. The equations of motion of the $A_\infty$ theory can be expressed in the form
\begin{equation}0=\widetilde{\M}\frac{1}{1-\widetilde{\Psi}},\label{eq:cohEOM}\end{equation}
where
\begin{equation}
\frac{1}{1-\widetilde{\Psi}}  = 1_{T\widetilde{\mathcal{H}}}\, +\, \widetilde{\Psi}\, +\widetilde{\Psi}\otimes\widetilde{\Psi} + \widetilde{\Psi}\otimes\widetilde{\Psi}\otimes\widetilde{\Psi} +...
\end{equation}
denotes the group-like element generated by $\widetilde{\Psi}$. Since
\begin{equation}\widetilde{\M} = \G^{-1}(\Q+\m_2|_2)\G,\end{equation}
multiplying \eq{cohEOM} by $\G$ gives
\begin{equation}
0 = (\Q+\m_2|_2)\G \frac{1}{1-\widetilde{\Psi}}.
\end{equation}
Let us look at the component of this equation with one NS output:
\begin{eqnarray}
0\lineup = \pi_1^0(\Q+\m_2|_2)\G \frac{1}{1-\widetilde{\Psi}}\nonumber\\
\lineup = Q\pi_1^0\G\frac{1}{1-\widetilde{\Psi}} + m_2\pi_2^2 \G\frac{1}{1-\widetilde{\Psi}}\nonumber\\
\lineup = Q\left(\pi_1^0\G\frac{1}{1-\widetilde{\Psi}}\right) + m_2\left(\pi_1^1 \G\frac{1}{1-\widetilde{\Psi}},\pi_1^1 \G\frac{1}{1-\widetilde{\Psi}}\right).\label{eq:BANSEOM1}
\end{eqnarray}
The component with one Ramond output is
\begin{eqnarray}
0\lineup = \pi_1^1(\Q+\m_2|_2)\G \frac{1}{1-\widetilde{\Psi}}\nonumber\\
\lineup = Q\pi_1^1\G\frac{1}{1-\widetilde{\Psi}} + m_2\pi_2^3 \G\frac{1}{1-\widetilde{\Psi}}\nonumber\\
\lineup = Q\left(\pi_1^1\G\frac{1}{1-\widetilde{\Psi}}\right).
\label{eq:BAREOM1}
\end{eqnarray}
Further note that 
\begin{eqnarray}
\pi_1^0\G\frac{1}{1-\widetilde{\Psi}}\lineup = \pi_1\G\frac{1}{1-\PsiNS},\\
\pi_1^1\G\frac{1}{1-\widetilde{\Psi}}\lineup = \pi_1\G\frac{1}{1-\PsiNS}\otimes\PsiR\otimes\frac{1}{1-\PsiNS},
\end{eqnarray}
and define
\begin{eqnarray}
A_\eta\lineup \equiv \pi_1\G\frac{1}{1-\PsiNS},\label{eq:An}\\
F\PsiR\lineup \equiv \pi_1\G\frac{1}{1-\PsiNS}\otimes\PsiR\otimes\frac{1}{1-\PsiNS}.\label{eq:FPsi}
\end{eqnarray}
Therefore \eq{BANSEOM1} and \eq{BAREOM1} reduce to
\begin{eqnarray}
0\lineup = QA_\eta +(F\PsiR)^2,\nonumber\\
0\lineup = QF\PsiR.
\end{eqnarray}
These are the same as the equations of motion of the WZW-based theory, \eq{lNSEOM} and \eq{lREOM}, with the ``hats" missing. It is therefore natural to suppose that the field redefinition between the theories is given by equating
\begin{eqnarray}
\widehat{A}_\eta \lineup = A_\eta,\\
\widehat{F}\hPsiR \lineup = F\PsiR.\label{eq:fieldred}
\end{eqnarray}
In the NS sector, this only specifies the field redefinition up to a gauge transformation of the form
\begin{equation}e^{\hPhiNS'} = e^{\hPhiNS} e^v,\ \ \ \ \eta v=0, \label{eq:4gt}\end{equation}
where $v$ is a gauge parameter, since this transformation leaves $\widehat{A}_\eta$ invariant. This ambiguity can be removed by partial gauge fixing \cite{INOT,OkWB,WB}, or by lifting the NS sector of the $A_\infty$ theory to the large Hilbert space \cite{WBlarge}, as will be reviewed in the next subsection.

To further simplify the field redefinition in the Ramond sector let us take a closer look at $F\PsiR$. Consider the expression:
\begin{equation}\pi_1^1 \G^{-1}\G\frac{1}{1-\PsiNS}\otimes\PsiR\otimes\frac{1}{1-\PsiNS}.\end{equation}
Canceling $\G^{-1}$ and $\G$ and projecting onto the 1-string output gives 
\begin{equation}\pi_1^1 \G^{-1}\G\frac{1}{1-\PsiNS}\otimes\PsiR\otimes\frac{1}{1-\PsiNS} = \PsiR.\label{eq:4dr1}\end{equation}
On the other hand, we can substitute \eq{GinvR} for $\pi_1^1\G^{-1}$, obtaining
\begin{eqnarray}
\lineup \pi_1^1 \G^{-1}\G \frac{1}{1-\PsiNS}\otimes\PsiR\otimes\frac{1}{1-\PsiNS} \nonumber\\
\lineup \ \ \ \  = \pi_1^1(\mathbb{I}_{T\widetilde{\mathcal{H}}} - \Xxit \m_2|_0 )\G\frac{1}{1-\PsiNS}\otimes\PsiR\otimes\frac{1}{1-\PsiNS}\nonumber\\
\lineup \ \ \ \ = \pi_1^1\G \frac{1}{1-\PsiNS}\otimes\PsiR\otimes\frac{1}{1-\PsiNS} - \Xit m_2|_0\pi_2^1\G \frac{1}{1-\PsiNS}\otimes\PsiR\otimes\frac{1}{1-\PsiNS}.\label{eq:4dr2}
\end{eqnarray}
The first term on the right hand side is $F\PsiR$. By writing
\begin{equation}
\pi_2^1 = \inverttriangle\!\Big[\pi_1^1\otimes'\pi_1^0+\pi_1^0\otimes'\pi_1^1\Big]\triangle
\end{equation}
we can show that the second term on the right hand side is
\begin{eqnarray}
\Xit m_2|_0 \pi_2^1\G \frac{1}{1-\PsiNS}\otimes\PsiR\otimes\frac{1}{1-\PsiNS} \lineup = 
 \Xit m_2|_0\big(F\PsiR\otimes A_\eta + A_\eta\otimes F\PsiR\big)\nonumber\\
 \lineup = \Xit [A_\eta,F\PsiR],\label{eq:OkDer1}
\end{eqnarray}
where in the last step we switched from degree to Grassmann grading.\footnote{The coproduct $\triangle$ acts on a group-like element as~\cite{WB}
\begin{equation}
\triangle \frac{1}{1-A} = \frac{1}{1-A} \otimes' \frac{1}{1-A}.
\end{equation}
A straightforward generalization gives the formulas 
\begin{eqnarray}
\triangle \frac{1}{1-A} \otimes B \otimes \frac{1}{1-A}
\lineup = 
\frac{1}{1-A} \otimes' \frac{1}{1-A} \otimes B \otimes \frac{1}{1-A}+\frac{1}{1-A} \otimes B \otimes \frac{1}{1-A} \otimes' \frac{1}{1-A},\ \ \ \ \ \ \ \ \\
\triangle \frac{1}{1-A} \otimes B \otimes \frac{1}{1-A} \otimes C \otimes \frac{1}{1-A}\lineup 
= \frac{1}{1-A} \otimes' \frac{1}{1-A} \otimes B \otimes \frac{1}{1-A} \otimes C \otimes \frac{1}{1-A}\nonumber\\
\lineup\ \ \ +\frac{1}{1-A} \otimes B \otimes \frac{1}{1-A} \otimes' \frac{1}{1-A} \otimes C \otimes \frac{1}{1-A}\nonumber\\
\lineup\ \ \ +\frac{1}{1-A} \otimes B \otimes \frac{1}{1-A} \otimes C \otimes \frac{1}{1-A} \otimes' \frac{1}{1-A}.
\end{eqnarray}
We use the first formula in the derivation of \eq{OkDer1}, and later the second formula in the derivation of~\eq{OkDer2} and the calculation of \eq{OkDer4} from \eq{OkDer3}.} Equating \eq{4dr1} and \eq{4dr2} then implies
\begin{equation}\PsiR = F\PsiR -\Xit[A_\eta,F\PsiR ].\end{equation}
This can be interpreted as a recursive formula for $F\PsiR$:
\begin{equation}
F\PsiR = \PsiR +\Xit [A_\eta,F\PsiR]. 
\end{equation}
Plugging this formula into itself implies
\begin{equation}
F\PsiR = \frac{1}{\mathbb{I}-\Xit\mathrm{ad}_{A_\eta}}\PsiR.
\end{equation}
This is the same formula which defines $\widehat{F}\hPsiR$, but with the ``hats" missing. Since the field redefinition in the NS sector implies $\widehat{A}_\eta=A_\eta$, the field redefinition in the Ramond sector simplifies to
\begin{equation}\hPsiR = \PsiR.\end{equation}
The Ramond fields are equal; there is no field redefinition between them. This was anticipated in~\cite{complete} and is not surprising for the following reason. Since the Ramond fields have identical kinetic terms, we can assume a field redefinition relating them takes the form
\begin{equation}\hPsiR = \PsiR + \mathscr{X}\Big(\widetilde{f}_2(\widetilde{\Psi},\widetilde{\Psi}) + \widetilde{f}_3(\widetilde{\Psi},\widetilde{\Psi},\widetilde{\Psi}) +...\Big),\end{equation}
where $\widetilde{f}_2,\widetilde{f}_3,...$ are string products and the factor of $\mathscr{X}$ is needed to ensure that both fields live in the restricted space. Since the interaction vertices of both theories are built out of $Q,\Xit$ and the open string star product, it is natural to assume that the field redefinition can be constructed from these operations. The $(n+2)$-product in the field redefinition $\widetilde{f}_{n+2}$ must carry ghost number $-n-1$. Therefore it must contain at least $n+1$ insertions of $\Xit$, since no other operations carry negative ghost number. This implies that $\widetilde{f}_{n+2}$ carries at least picture $n+1$, and $\widetilde{f}_{n+2}(\widetilde{\Psi},...,\widetilde{\Psi})$ must have picture greater than or equal to $-1$. However, consistency of the field redefinition requires that  $\widetilde{f}_{n+2}(\widetilde{\Psi},...,\widetilde{\Psi})$ carries picture $-3/2$. Therefore $\widetilde{f}_{n+2}$ must vanish, and the Ramond fields are equal.

We therefore conclude that the field redefinition between the $A_\infty$ theory and WZW-based theory is
\begin{eqnarray}
\widehat{A}_\eta \lineup = A_\eta,\label{eq:fieldred2}\\
\hPsiR \lineup = \PsiR,\label{eq:fieldred3}
\end{eqnarray}
up to a gauge transformation of the form \eq{4gt}. It is important to note that the proposed field redefinition is consistent with the assumption that $\PsiNS$ and $\PsiR$ are in the small Hilbert space. In the Ramond sector this is obvious. In the NS sector it follows from the fact that $A_\eta$ and $\widehat{A}_\eta$ satisfy
\begin{equation}
\eta A_\eta - A_\eta*A_\eta = 0,\ \ \ \eta \widehat{A}_\eta - \widehat{A}_\eta*\widehat{A}_\eta = 0. 
\end{equation}
See \cite{OkWB,WB}. 

\subsection{Equivalence of the Actions}

Here we demonstrate that the field redefinition given by \eq{fieldred2} and \eq{fieldred3} relates the theories at the level of the action, not just the equations of motion. Following the analysis of \cite{OkWB,WBlarge}, this can be demonstrated by expressing the $A_\infty$ action in the same form as the WZW-based action, including the contribution from the Ramond sector. Let us explain how this is done.

The $(n+3)$-string vertex in the $A_\infty$ action is 
\begin{equation}\frac{1}{n+3}\widetilde{\omega}(\widetilde{\Psi},\widetilde{M}_{n+2}(\widetilde{\Psi},...,\widetilde{\Psi})).\end{equation}
Let us expand $\widetilde{\Psi}$ into NS and R components. Since the composite products multiply at most two Ramond states, the expanded vertex takes the form
\begin{eqnarray}
\lineup \frac{1}{n+3}\widetilde{\omega}(\widetilde{\Psi},\widetilde{M}_{n+2}(\widetilde{\Psi},...,\widetilde{\Psi}))=
\frac{1}{n+3}\Bigg[\widetilde{\omega}(\PsiNS,\widetilde{M}_{n+2}(\PsiNS,...,\PsiNS))\nonumber\\
\lineup\ \ \ \ \ \ \ \ \ \ \ \ \ \ \ +\sum_{k=0}^{n+1} \widetilde{\omega}(\PsiR,\widetilde{M}_{n+2}(\underbrace{\PsiNS,...,
\PsiNS}_{k\ \mathrm{times}},\PsiR,\PsiNS,...,\PsiNS))\nonumber\\
\lineup\ \ \ \ \ \ \ \ \ \ \ \ \ \ \ +\sum_{k=0}^n\sum_{j=0}^{n-k}\widetilde{\omega}(\PsiNS,\widetilde{M}_{n+2}(\underbrace{\PsiNS,...,\PsiNS}_{k\ \mathrm{times}},\PsiR,
\underbrace{\PsiNS,...,\PsiNS}_{j\ \mathrm{times}},\PsiR,\PsiNS,...,\PsiNS))\Bigg].\ \ \ \ \ \ \ \ \ \ 
\end{eqnarray}
Many terms in these sums are redundant. In fact, using cyclicity we can write the sum in the second line as $2/(n+1)$ times the double sum in the third line. Therefore we have
\begin{eqnarray}
\lineup \frac{1}{n+3}\widetilde{\omega}(\widetilde{\Psi},\widetilde{M}_{n+2}(\widetilde{\Psi},...,\widetilde{\Psi}))=
\frac{1}{n+3}\widetilde{\omega}(\PsiNS,\widetilde{M}_{n+2}(\PsiNS,...,\PsiNS))\nonumber\\
\lineup\ \ \ \ \ \ \ \ \ \ \ \ \ \ \ +\frac{1}{n+1}\sum_{k=0}^n\sum_{j=0}^{n-k}\widetilde{\omega}(\PsiNS,\widetilde{M}_{n+2}(\underbrace{\PsiNS,...,\PsiNS}_{k\ \mathrm{times}},\PsiR,
\underbrace{\PsiNS,...,\PsiNS}_{j\ \mathrm{times}},\PsiR,\PsiNS,...,\PsiNS)).\ \ \ \ \ \ \ \ 
\end{eqnarray}
Next we introduce a one-parameter family of NS string fields $\PsiNS(t),t\in[0,1]$ subject to the boundary conditions
\begin{equation}\PsiNS(0) = 0,\ \ \ \ \PsiNS(1) =\PsiNS.\end{equation}
The $(n+3)$-string vertex can be written as the integral of a total derivative in $t$:
\begin{eqnarray}
\lineup\!\!\!\!\!\!\!\!\!\! \frac{1}{n+3}\widetilde{\omega}(\widetilde{\Psi},\widetilde{M}_{n+2}(\widetilde{\Psi},...,\widetilde{\Psi}))=
\int_0^1 dt\,\frac{d}{dt}\bigg[\frac{1}{n+3}\widetilde{\omega}(\PsiNS(t),\widetilde{M}_{n+2}(\PsiNS(t),...,\PsiNS(t)))\nonumber\\
\lineup\!\!\!\!\!\!\!\!\!\!
+\frac{1}{n+1}\!\sum_{k=0}^n\sum_{j=0}^{n-k}\widetilde{\omega}(\PsiNS(t),\widetilde{M}_{n+2}(\underbrace{\PsiNS(t),...,\PsiNS(t)}_{k\ \mathrm{times}},\PsiR,
\underbrace{\PsiNS(t),...,\PsiNS(t)}_{j\ \mathrm{times}},\PsiR,\PsiNS(t),...,\PsiNS(t)))\!\bigg].\nonumber\\
\end{eqnarray}
Acting $d/dt$ produces $n+3$ terms with $\dot{\Psi}_\mathrm{NS}(t)=d\PsiNS(t)/dt$ in the first line, and in the second term it produces $n+1$ terms with $\dot{\Psi}_\mathrm{NS}(t)$. All of these terms are related by cyclicity, and therefore we can bring $\dot{\Psi}_\mathrm{NS}(t)$ to the first entry of the symplectic form and cancel the factors $1/(n+3)$ and $1/(n+1)$:
\begin{eqnarray}
\lineup\!\!\!\!\!\!\!\!\!\! \frac{1}{n+3}\widetilde{\omega}(\widetilde{\Psi},\widetilde{M}_{n+2}(\widetilde{\Psi},...,\widetilde{\Psi}))=
\int_0^1 dt \bigg[\omega_S(\dot{\Psi}_{\mathrm{NS}}(t),\widetilde{M}_{n+2}(\PsiNS(t),...,\PsiNS(t)))\nonumber\\
\lineup\!\!\!\!\!\!\!\!\!\!
+\sum_{k=0}^n\sum_{j=0}^{n-k}\omega_S(\dot{\Psi}_\mathrm{NS}(t),\widetilde{M}_{n+2}(\underbrace{\PsiNS(t),...,\PsiNS(t)}_{k\ \mathrm{times}},\PsiR,
\underbrace{\PsiNS(t),...,\PsiNS(t)}_{j\ \mathrm{times}},\PsiR,\PsiNS(t),...,\PsiNS(t)))\bigg].\nonumber\\
\end{eqnarray}
On the right hand side we replaced $\widetilde{\omega}$ with $\omega_S$ since only NS states are contracted. We can simplify this expression using coderivations and group-like elements:
\begin{eqnarray}
\lineup\frac{1}{n+3}\widetilde{\omega}(\widetilde{\Psi},\widetilde{M}_{n+2}(\widetilde{\Psi},...,\widetilde{\Psi}))=
\int_0^1 dt \left[\omega_S\left(\dot{\Psi}_{\mathrm{NS}}(t),\pi_1\M_{n+2}|_0\frac{1}{1-\PsiNS(t)}\right)\right.\nonumber\\
\lineup\ \ \ \ \ \ \ \ \ \ 
+\left.\omega_S\left(\dot{\Psi}_\mathrm{NS}(t),\pi_1\m_{n+2}|_2\frac{1}{1-\PsiNS(t)}\otimes \PsiR\otimes \frac{1}{1-\PsiNS(t)}\otimes\PsiR\otimes \frac{1}{1-\PsiNS(t)}\right)\right].\ \ \ \ \ \ \ \ \ \ \ 
\end{eqnarray}
Summing over the vertices, the action can therefore be expressed as
\begin{eqnarray}
S \lineup = \frac{1}{2}\widetilde{\omega}(\widetilde{\Psi},Q\widetilde{\Psi}) +\sum_{n=0}^\infty \frac{1}{n+3} \widetilde{\omega}(\widetilde{\Psi},\widetilde{M}_{n+2}(\widetilde{\Psi},...\widetilde{\Psi}))\nonumber\\
\lineup = \frac{1}{2}\omega_S(\mathscr{Y}\PsiR,Q\PsiR)+ \frac{1}{2}\omega_S(\PsiNS,Q\PsiNS) + 
\int_0^1 dt \left[\omega_S\!\!\left(\dot{\Psi}_{\mathrm{NS}}(t),\pi_1(\M|_0-\Q)\frac{1}{1-\PsiNS(t)}\right)\right.\nonumber\\
\lineup\ \ \ \ \ \ \ \ \ \ 
+\left.\omega_S\left(\dot{\Psi}_\mathrm{NS}(t),\pi_1\m|_2\frac{1}{1-\PsiNS(t)}\otimes \PsiR\otimes \frac{1}{1-\PsiNS(t)}\otimes\PsiR\otimes \frac{1}{1-\PsiNS(t)}\right)\right].
\end{eqnarray}
We can absorb the NS kinetic term into the integral over $t$, obtaining
\begin{eqnarray}
S\lineup = \frac{1}{2}\omega_S(\mathscr{Y}\PsiR,Q\PsiR) + 
\int_0^1 dt \left[\omega_S\left(\dot{\Psi}_{\mathrm{NS}}(t),\pi_1\M|_0\frac{1}{1-\PsiNS(t)}\right)\right.\nonumber\\
\lineup\ \ \ \ \ \ \ \ \ \ 
+\left.\omega_S\left(\dot{\Psi}_\mathrm{NS}(t),\pi_1\m|_2\frac{1}{1-\PsiNS(t)}\otimes \PsiR\otimes \frac{1}{1-\PsiNS(t)}\otimes\PsiR\otimes \frac{1}{1-\PsiNS(t)}\right)\right].\ \ \ \ \ \ \ \ \ \ \ 
\end{eqnarray}
Because this form of the action was constructed from the integral of a total derivative, it only depends on the value of $\PsiNS(t)$ at $t=1$.

Next it will be helpful to reformulate the theory in the large Hilbert space. We replace $\PsiNS$ with a new NS string field $\PhiNS$ in the large Hilbert space according to
\begin{equation}\PsiNS = \eta\PhiNS.\end{equation}
The new field $\PhiNS$ is degree odd (because it is Grassmann even) and carries ghost and picture number zero. We also introduce a corresponding family of string fields $\PhiNS(t),t\in[0,1]$ such that $\eta\PhiNS(t) = \PsiNS(t)$. Plugging into the action gives
\begin{eqnarray}
S\lineup = \frac{1}{2}\omega_S(\mathscr{Y}\PsiR,Q\PsiR) + 
\int_0^1 dt \left[\omega_L\left(\dot{\Phi}_{\mathrm{NS}}(t),\pi_1\M|_0\frac{1}{1-\eta\PhiNS(t)}\right)\right.\nonumber\\
\lineup\ \ \ \ \ \ \ \ \ \ 
+\left.\omega_L\left(\dot{\Phi}_\mathrm{NS}(t),\pi_1\m|_2\frac{1}{1-\eta\PhiNS(t)}\otimes \PsiR\otimes \frac{1}{1-\eta\PhiNS(t)}\otimes\PsiR\otimes \frac{1}{1-\eta\PhiNS(t)}\right)\right].\ \ \ \ \ \ \ \ \ \ \ \label{eq:BAact1}
\end{eqnarray}
Here we replaced the small Hilbert space symplectic form with the large Hilbert space symplectic form using the relation
\begin{equation}
\omega_S(\eta \Phi,\Psi)=\omega_L(\Phi,\Psi),
\end{equation}
where $\Phi$ is in the large Hilbert space and $\Psi$ is in the small Hilbert space. Next we use the identity~\cite{OkWB,WBlarge}
\begin{equation}
\omega(B,C)=\omega\left(\pi_1\H\frac{1}{1-A}\otimes B\otimes\frac{1}{1-A},\pi_1\H\frac{1}{1-A}\otimes C\otimes\frac{1}{1-A}\right),
\end{equation}
where $B$ and $C$ are string fields, $A$ is a degree even string field, and the cohomomorphism $\H$ is cyclic with respect to $\omega$. In the current application
we identify
\begin{equation}A\to \eta\PhiNS(t),\ \ \ \H\to\G,\ \ \ \omega\to\omega_L. \end{equation}
Note, in particular, that $\G$ is cyclic with respect to the large Hilbert space symplectic form when it receives no Ramond inputs. Thus we can rewrite the action as follows: 
\begin{eqnarray}
S \lineup = \frac{1}{2}\omega_S(\mathscr{Y}\PsiR,Q\PsiR)\nonumber\\
\lineup \ \ \ + \int_0^1 dt\,\omega_L\left(\pi_1\G\frac{1}{1-\eta\PhiNS(t)}\otimes \dot{\Phi}_{\mathrm{NS}}(t)\otimes \frac{1}{1-\eta\PhiNS(t)},\right.\nonumber\\
\lineup\ \ \ \ \ \ \ \ \ \ \ \ \ \ \ \ \ \ \ \ \ \ \ \ \ \ \ \ \ \ \ \ \ \ \ \ \ \ \ \ \ \ 
\left.\pi_1\G\frac{1}{1-\eta\PhiNS(t)}\otimes\left(\pi_1\M|_0\frac{1}{1-\eta\PhiNS(t)}\right)\otimes\frac{1}{1-\eta\PhiNS(t)}\right)
\nonumber\\
\lineup\ \ \ 
+ \int_0^1 dt\,\omega_L\left(\pi_1\G\frac{1}{1-\eta\PhiNS(t)}\otimes \dot{\Phi}_{\mathrm{NS}}(t)\otimes \frac{1}{1-\eta\PhiNS(t)},\pi_1\G\frac{1}{1\!-\!\eta\PhiNS(t)} \otimes \right.\nonumber\\
\lineup\ \ \ \ \ \ \ \ \ \ \ \ \ \ \ \ \ \ \ \ \ \ \ \left.\left(\pi_1\m|_2\frac{1}{1\!-\!\eta\PhiNS(t)}\!\otimes\! \PsiR\!\otimes\! \frac{1}{1\!-\!\eta\PhiNS(t)}\!\otimes\!\PsiR\!\otimes\! \frac{1}{1\!-\!\eta\PhiNS(t)}\right)\!\otimes\!\frac{1}{1\!-\!\eta\PhiNS(t)}\right).\nonumber\\
\end{eqnarray}
We can simplify the term with $\M|_0$ by writing
\begin{equation}
\frac{1}{1-\eta\PhiNS(t)}\otimes\left(\pi_1\M|_0\frac{1}{1-\eta\PhiNS(t)}\right)\otimes\frac{1}{1-\eta\PhiNS(t)}=\M|_0\frac{1}{1-\eta\PhiNS(t)}.
\end{equation}
The term with $\m|_2$ can also be simplified using 
\begin{eqnarray}
\lineup \frac{1}{1\!-\!\eta\PhiNS(t)}\!\otimes\!\left(\pi_1\m|_2\frac{1}{1\!-\!\eta\PhiNS(t)}\!\otimes\! \PsiR\!\otimes\! \frac{1}{1\!-\!\eta\PhiNS(t)}\!\otimes\!\PsiR\!\otimes\! \frac{1}{1\!-\!\eta\PhiNS(t)}\right)\!\otimes\!\frac{1}{1\!-\!\eta\PhiNS(t)}\nonumber\\
\lineup\ \ \ \ \  = \m|_2\frac{1}{1\!-\!\eta\PhiNS(t)}\!\otimes\! \PsiR\!\otimes\! \frac{1}{1\!-\!\eta\PhiNS(t)}\!\otimes\!\PsiR\!\otimes\! \frac{1}{1\!-\!\eta\PhiNS(t)}.
\end{eqnarray}
Therefore, we have
\begin{eqnarray}
S \lineup = \frac{1}{2}\omega_S(\mathscr{Y}\PsiR,Q\PsiR)\nonumber\\
\lineup\ \ \ + \int_0^1 dt\,\omega_L\left(\pi_1\G\frac{1}{1-\eta\PhiNS(t)}\otimes \dot{\Phi}_{\mathrm{NS}}(t)\otimes \frac{1}{1-\eta\PhiNS(t)},
\pi_1\G\M|_0\frac{1}{1-\eta\PhiNS(t)}\right)
\nonumber\\
\lineup\ \ \ 
+ \int_0^1 dt\,\omega_L\!\left(\pi_1\G\frac{1}{1\!-\!\eta\PhiNS(t)}\!\otimes\! \dot{\Phi}_{\mathrm{NS}}(t)\!\otimes\! \frac{1}{1\!-\!\eta\PhiNS(t)},\right.\nonumber\\
\lineup\ \ \ \ \ \ \ \ \ \ \ \ \ \ \ \ \ \ \ \ \ \ \ 
\left.\pi_1\G\m|_2\frac{1}{1\!-\!\eta\PhiNS(t)}\!\otimes\! \PsiR\!\otimes\! \frac{1}{1\!-\!\eta\PhiNS(t)}\!\otimes\!\PsiR\!\otimes\! \frac{1}{1\!-\!\eta\PhiNS(t)}\right).\nonumber\\
\end{eqnarray}
Now using 
\begin{eqnarray}
\pi_1\G\M|_0 \lineup = \pi_1\Q\G = Q\pi_1\G,\\
\pi_1\G\m|_2 \lineup = \pi_1\m_2|_2\G = m_2\pi_2^2\G,
\end{eqnarray}
we further obtain
\begin{eqnarray}
S \lineup = \frac{1}{2}\omega_S(\mathscr{Y}\PsiR,Q\PsiR)\nonumber\\
\lineup\ \ \ + \int_0^1 dt\,\omega_L\left(\pi_1\G\frac{1}{1-\eta\PhiNS(t)}\otimes \dot{\Phi}_{\mathrm{NS}}(t)\otimes \frac{1}{1-\eta\PhiNS(t)},
Q\pi_1\G\frac{1}{1-\eta\PhiNS(t)}\right)
\nonumber\\
\lineup\ \ \ 
+ \int_0^1 dt\,\omega_L\!\left(\pi_1\G\frac{1}{1\!-\!\eta\PhiNS(t)}\!\otimes\! \dot{\Phi}_{\mathrm{NS}}(t)\!\otimes\! \frac{1}{1\!-\!\eta\PhiNS(t)},\right.\nonumber\\
\lineup\ \ \ \ \ \ \ \ \ \ \ \ \ \ \ \ \ \ \ \ \ \ \ 
\left.m_2\pi_2^2\G\frac{1}{1\!-\!\eta\PhiNS(t)}\!\otimes\! \PsiR\!\otimes\! \frac{1}{1\!-\!\eta\PhiNS(t)}\!\otimes\!\PsiR\!\otimes\! \frac{1}{1\!-\!\eta\PhiNS(t)}\right).
\end{eqnarray}
Using $\pi_2^2 = \inverttriangle\![\pi_1^1\otimes'\pi_1^1]\triangle$, one can show that 
\begin{eqnarray}
\lineup \pi_2^2\G\frac{1}{1\!-\!\eta\PhiNS(t)}\!\otimes\! \PsiR\!\otimes\! \frac{1}{1\!-\!\eta\PhiNS(t)}\!\otimes\!\PsiR\!\otimes\! \frac{1}{1\!-\!\eta\PhiNS(t)} = \Big(F(t)\PsiR\Big) \otimes \Big(F(t)\PsiR\Big),\label{eq:OkDer2}
\end{eqnarray}
where 
\begin{equation}
F(t)\PsiR \equiv \pi_1\G\frac{1}{1\!-\!\eta\PhiNS(t)}\!\otimes\! \PsiR\!\otimes\! \frac{1}{1\!-\!\eta\PhiNS(t)}.
\end{equation}
Switching from degree to Grassmann grading, the action is therefore expressed as 
\begin{equation}
S = \frac{1}{2}\langle\mathscr{Y}\PsiR,Q\PsiR\rangle_S- \int_0^1 dt\,\langle A_t(t), QA_\eta(t) + (F(t)\PsiR)^2\rangle_L,\label{eq:sHSaction}
\end{equation}
where following \cite{OkWB,WBlarge} we define the potentials by
\begin{eqnarray}
A_t(t)\lineup \equiv \pi_1\G\frac{1}{1-\eta\PhiNS(t)}\otimes \dot{\Phi}_{\mathrm{NS}}(t)\otimes \frac{1}{1-\eta\PhiNS(t)},\\
A_\eta(t) \lineup \equiv \pi_1\G \frac{1}{1-\eta\PhiNS(t)}.
\end{eqnarray}
Thus the $A_\infty$ action is expressed in the same form as \eq{lHSaction} but with the ``hats" missing.

Now we can show that the action of the $A_\infty$ theory is related to the action of the WZW-based theory by field redefinition. We postulate that the two theories are related by 
\begin{equation}
\widehat{A}_t(t) = A_t(t),\ \ \ \ \hPsiR = \PsiR.\label{eq:tid}
\end{equation} 
Equating the $t$-potentials provides an invertible map between $\PhiNS(t)$ and $\hPhiNS(t)$, and automatically equates the $\eta$-potentials \cite{WBlarge}:
\begin{equation}\widehat{A}_\eta(t) = A_\eta(t).\end{equation} 
With these identifications it is identically true that the actions \eq{lHSaction} and \eq{sHSaction} are equal. Moreover, since the $A_\infty$ action is only a function of $\PsiNS(t) = \eta\PhiNS(t)$ at $t=1$, the identification \eq{tid} is equivalent to 
\begin{equation}
\widehat{A}_\eta = A_\eta,\ \ \ \ \hPsiR = \PsiR,
\end{equation} 
which is the field redefinition anticipated in the previous subsection.

\section{Conclusions}

In this paper we have constructed the NS and R sectors of open superstring field theory realizing a cyclic $A_\infty$ structure. This means, in particular, that we have an explicit solution of the classical Batalin-Vilkovisky master equation,
\begin{equation}
\{S,S\} = 0,
\end{equation}
after relaxing the ghost number constraint on the NS and R string fields. Therefore, for the purpose of tree level amplitudes we have a consistent definition of the gauge-fixed path integral, and for the first time we are prepared to consider quantum effects in superstring field theory. 

However, the absence of explicit closed string fields and the appearance of spurious singularities at higher genus may make quantization subtle. Therefore it is desirable to give a construction of superstring field theory realizing a more general decomposition of the bosonic moduli space than is provided by the Witten vertex. This in turn is closely related to the generalization to heterotic and type II closed superstring field theories. The appropriate construction of NS actions and Ramond equations of motion is described in \cite{ClosedSS,Ramond}, and in principle all that is needed is to implement cyclicity. For example, in the closely related open string field theory with stubs \cite{ClosedSS,Ramond}, it is not difficult to see that the gauge products with one Ramond output and zero picture deficit should be defined by
\begin{equation}\mu_{n+2}^{(n-r+1)}|_{2r}=\Xit M_{n+2}^{(n-r)}|_{2r}\ \ \ \ (2r+1\ \mathrm{Ramond\ inputs}),\end{equation}
so that the equations of motion are consistent with the projection onto the restricted space in the Ramond sector. However, a full specification of the vertices requires many additional gauge products of varying Ramond numbers and picture deficits. Solving the entire recursive system of products consistent with cyclicity is a much more challenging problem, which we hope to consider soon.

\vspace{.5cm}

\noindent{\bf Acknowledgments}

\vspace{.25cm}

\noindent T.E. would like to thank S. Konopka and I. Sachs for discussion. The work of T.E. was supported in part by the DFG Transregional Collaborative Research Centre TRR 33 and the DFG cluster of excellence Origin and Structure of the Universe. The work of Y.O. was supported in part by a Grant-in-Aid for Scientific Research (B) No.~25287049 and a Grant-in-Aid for Scientific Research~(C) No.~24540254 from the Japan Society for the Promotion of Science (JSPS).

\begin{appendix}

\section{Gauge Invariance in Sen's Formulation}
\label{app:Sen}

In Sen's formulation of the Ramond sector \cite{1PIR,SenBV}, the action does not realize a cyclic $A_\infty$ structure in the standard sense. Therefore it is worth explaining why it is gauge invariant. The infinitesimal gauge transformation can be written in the form
\begin{eqnarray}
\delta\widetilde{\Pi} \lineup = Q\widetilde{\Omega} + \pi_1\widetilde{\b}\frac{1}{1-\widetilde{\Psi}}\otimes \widetilde{\Lambda}\otimes\frac{1}{1-\widetilde{\Psi}},\\
\delta\widetilde{\Psi} \lineup = \pi_1\widetilde{\M}\frac{1}{1-\widetilde{\Psi}}\otimes \widetilde{\Lambda}\otimes\frac{1}{1-\widetilde{\Psi}},
\end{eqnarray}
where $\widetilde{\Omega}$ and $\widetilde{\Lambda}$ are degree odd gauge parameters in the small Hilbert space, at ghost number zero, and with the appropriate picture in the NS and R sectors. The variation of the action is
\begin{equation}
\delta S = -\omega_S(\delta\widetilde{\Pi},\mathcal{G}Q\widetilde{\Pi})  + \omega_S(\delta\widetilde{\Psi},Q\widetilde{\Pi})+ \omega_S(\delta\widetilde{\Pi},Q\widetilde{\Psi})+\omega_S\left(\delta\widetilde{\Psi},\pi_1\widetilde{\b}\frac{1}{1-\widetilde{\Psi}}\right).
\end{equation}
The gauge parameter $\widetilde{\Omega}$ immediately drops out since $Q\widetilde{\Omega}$ always appears in the symplectic form contracted with a BRST invariant state. Substituting the infinitesimal gauge transformation then gives
\begin{eqnarray}
\delta S \lineup = -\omega_S\left(\pi_1\widetilde{\b}\frac{1}{1-\widetilde{\Psi}}\otimes\widetilde{\Lambda}\otimes\frac{1}{1-\widetilde{\Psi}},\mathcal{G}Q\widetilde{\Pi}\right)  + \omega_S\left(\pi_1\widetilde{\M}\frac{1}{1-\widetilde{\Psi}}\otimes\widetilde{\Lambda}\otimes\frac{1}{1-\widetilde{\Psi}},Q\widetilde{\Pi}\right)\nonumber\\
\lineup \ \ \ + \omega_S\left(\pi_1\widetilde{\b}\frac{1}{1-\widetilde{\Psi}}\otimes\widetilde{\Lambda}\otimes\frac{1}{1-\widetilde{\Psi}},Q\widetilde{\Psi}\right)+\omega_S\left(\pi_1\widetilde{\M}\frac{1}{1-\widetilde{\Psi}}\otimes\widetilde{\Lambda}\otimes\frac{1}{1-\widetilde{\Psi}},\pi_1\widetilde{\b}\frac{1}{1-\widetilde{\Psi}}\right).\ \ \ \ 
\end{eqnarray}
The first and second terms cancel upon using the BPZ even property of $\mathcal{G}$ and converting $\widetilde{\b}$ into $\widetilde{\M}-\Q$. In the last term  we replace $\pi_1\widetilde{\M}$ with $\pi_1\Q+\mathcal{G}\pi_1\widetilde{\b}$:
\begin{equation}
\delta S  = \omega_S\left(\pi_1\widetilde{\b}\frac{1}{1-\widetilde{\Psi}}\otimes\widetilde{\Lambda}\otimes\frac{1}{1-\widetilde{\Psi}},Q\widetilde{\Psi}\right)+\omega_S\left(Q\widetilde{\Lambda} + \mathcal{G}\pi_1\widetilde{\b}\frac{1}{1-\widetilde{\Psi}}\otimes\widetilde{\Lambda}\otimes\frac{1}{1-\widetilde{\Psi}},\pi_1\widetilde{\b}\frac{1}{1-\widetilde{\Psi}}\right).
\end{equation}
Next use the BPZ even property of $\mathcal{G}$ and again convert $\widetilde{\b}$ into $\widetilde{\M}-\Q$:
\begin{eqnarray}
\delta S\lineup = \omega_S\left(\pi_1\widetilde{\b}\frac{1}{1-\widetilde{\Psi}}\otimes\widetilde{\Lambda}\otimes\frac{1}{1-\widetilde{\Psi}},Q\widetilde{\Psi}\right)+\omega_S\left(Q\widetilde{\Lambda},\pi_1\widetilde{\b}\frac{1}{1-\widetilde{\Psi}}\right) \nonumber\\
\lineup\ \ \ + \omega_S\left(\pi_1\widetilde{\b}\frac{1}{1-\widetilde{\Psi}}\otimes\widetilde{\Lambda}\otimes\frac{1}{1-\widetilde{\Psi}},\pi_1(\widetilde{\M}-\Q)\frac{1}{1-\widetilde{\Psi}}\right)\nonumber\\
\lineup = \omega_S\left(\widetilde{\Lambda},\pi_1\Q\widetilde{\b}\frac{1}{1-\widetilde{\Psi}}\right)+ \omega_S\left(\pi_1\widetilde{\b}\frac{1}{1-\widetilde{\Psi}}\otimes\widetilde{\Lambda}\otimes\frac{1}{1-\widetilde{\Psi}},\pi_1\widetilde{\M}\frac{1}{1-\widetilde{\Psi}}\right).\label{eq:Sengv1}
\end{eqnarray}
Using cyclicity of $\widetilde{\b}$ we can rewrite the second term as
\begin{equation}
\omega_S\left(\pi_1\widetilde{\b}\frac{1}{1-\widetilde{\Psi}}\otimes\widetilde{\Lambda}\otimes\frac{1}{1-\widetilde{\Psi}},\pi_1\widetilde{\M}\frac{1}{1-\widetilde{\Psi}}\right)= \omega_S\left(\widetilde{\Lambda},\pi_1\widetilde{\b}\frac{1}{1-\widetilde{\Psi}}\otimes\left(\pi_1\widetilde{\M}\frac{1}{1-\widetilde{\Psi}}\right)\otimes \frac{1}{1-\widetilde{\Psi}}\right).\label{eq:OkDer4}
\end{equation}
This follows from the relation 
\begin{equation}
0 = \langle \omega_S| \pi_2\widetilde{\b}\frac{1}{1-\widetilde{\Psi}}\otimes\widetilde{\Lambda}\otimes\frac{1}{1-\widetilde{\Psi}}\otimes\left(\pi_1\widetilde{\M}\frac{1}{1-\widetilde{\Psi}}\right)\otimes\frac{1}{1-\widetilde{\Psi}},\label{eq:OkDer3}
\end{equation}
after representing $\pi_2 = \inverttriangle\!\Big[\pi_1\otimes'\pi_1\Big]\triangle$ and acting with the coproduct. Therefore the gauge variation of the action produces 
\begin{eqnarray}
\delta S\lineup = \omega_S\left(\widetilde{\Lambda},\pi_1\Q\widetilde{\b}\frac{1}{1-\widetilde{\Psi}}\right)+
\omega_S\left(\widetilde{\Lambda},\pi_1\widetilde{\b}\frac{1}{1-\widetilde{\Psi}}\otimes\left(\pi_1\widetilde{\M}\frac{1}{1-\widetilde{\Psi}}\right)\otimes\frac{1}{1-\widetilde{\Psi}}\right)\nonumber\\
\lineup = \omega_S\left(\widetilde{\Lambda},\pi_1(\Q\widetilde{\b} + \widetilde{\b}\widetilde{\M})\frac{1}{1-\widetilde{\Psi}}\right)\nonumber\\
\lineup = 0,
\end{eqnarray}
which vanishes as a consequence of \eq{bcid}.

\end{appendix}

\end{document}